\pdfoutput=1
\documentclass{emulateapj}
\usepackage{color}

\shorttitle{Orphan Flares \& Jet Sheaths}
\shortauthors{MacDonald, Jorstad, \& Marscher}

\begin{document}

\title{``Orphan'' $\gamma$-ray Flares and Stationary Sheaths of Blazar Jets}

\author{Nicholas R. MacDonald\altaffilmark{1,2}, Svetlana G. Jorstad\altaffilmark{1,3}, and Alan P. Marscher\altaffilmark{1}}
\affil{$~^{1}$Institute for Astrophysical Research, Boston University, 725 Commonwealth Avenue, Boston, MA 02215}
\affil{$~^{2}$Max-Planck-Institut f\"{u}r Radioastronomie, Auf dem H\"{u}gel 69, D- 53121 Bonn, Germany}
\affil{$~^{3}$Astronomical Institute, St. Petersburg State University, Universitetskij Pr. 28, Petrodvorets, 198504 St. Petersburg, Russia}

\begin{abstract}
Blazars exhibit flares across the entire electromagnetic spectrum. Many $\gamma$-ray flares are highly correlated with flares detected at longer wavelengths; however, a small subset appears to occur in isolation, with little or no correlated variability at longer wavelengths.  These ``orphan'' $\gamma$-ray flares challenge current models of blazar variability, most of which are unable to reproduce this type of behavior.  \citeauthor{macdonald15} have developed the \textit{Ring of Fire} model to explain the origin of orphan $\gamma$-ray flares from within blazar jets.  In this model, electrons contained within a blob of plasma moving relativistically along the spine of the jet inverse-Compton scatter synchrotron photons emanating off of a ring of shocked sheath plasma that enshrouds the jet spine.  As the blob propagates through the ring, the scattering of the ring photons by the blob electrons creates an orphan $\gamma$-ray flare.  This model was successfully applied to modeling a prominent orphan $\gamma$-ray flare observed in the blazar PKS 1510$-$089.  To further support the plausibility of this model, \citeauthor{macdonald15} presented a stacked radio map of PKS 1510$-$089 containing the polarimetric signature of a sheath of plasma surrounding the spine of the jet.  In this paper, we extend our modeling and stacking techniques to a larger sample of blazars: 3C 273, 4C 71$.$01, 3C 279, 1055$+$018, CTA 102, and 3C 345, the majority of which have exhibited orphan $\gamma$-ray flares.  We find that the model can successfully reproduce these flares, while our stacked maps reveal the existence of jet sheaths within these blazars.   
\end{abstract}


\section{Introduction}

Many supermassive black hole systems at the centers of galaxies produce relativistic jets that propagate from sub-parsec to kilo-parsec scales.  These radio jets emit synchrotron radiation due to the presence of electrons gyrating about magnetic field lines within the jet.  A sub-set of these radio jets are closely aligned to our line-of-sight.  These aligned jets are referred to as ``blazars''.  At 43 GHz, a VLBI image of a blazar typically exhibits a bright stationary feature known as the ``radio core''.  It has been postulated that the radio core of a blazar imaged at millimeter wavelengths is associated with a standing recollimation shock (resulting from pressure imbalance between the jet and its surroundings) several parsecs downstream of the central super-massive black hole of the system (see \citealt{marscher08}; \citealt{cawthorne13}; \citealt{dodson16} and references therein).  Over the course of time, ``blobs'' of plasma are ejected from the radio core and propagate down the jet at relativistic speeds, which can appear superluminal in the observer's frame.  These blobs could be either internal shock waves propagating through the jet (see, e.g., \citealt{joshi11}) or plasmoids with higher density and/or magnetic field than the ambient flow (the view adopted in this paper).  

Blazars emit light across the entire electromagnetic spectrum, from low-energy radio waves to high-energy $\gamma$-rays.  With the launch of the Fermi Large Area Telescope (LAT), it has become apparent (as hinted by EGRET) that the $\gamma$-ray sky outside the plane of the Milky Way is dominated by blazar emission.  Blazars are highly variable with timescales of $\gamma$-ray variability ranging from months, to days, and even minutes (e.g., \citealt{aharonian07}).  Prominent $\gamma$-ray flares tend to be correlated with flares observed at longer wavelengths (e.g., in the optical and the radio; \citealt{marscher12}).  There is, however, a class of ``orphan'' $\gamma$-ray flares that occur in apparent isolation, with little correlated variability detected at longer wavelengths.  Standard shock models of blazar emission have difficulty reproducing these orphan $\gamma$-ray flares.  

\begin{figure*}
  \setlength{\abovecaptionskip}{-6pt}
  \begin{center}
    \scalebox{0.87}{\includegraphics[width=2.0\columnwidth,clip]{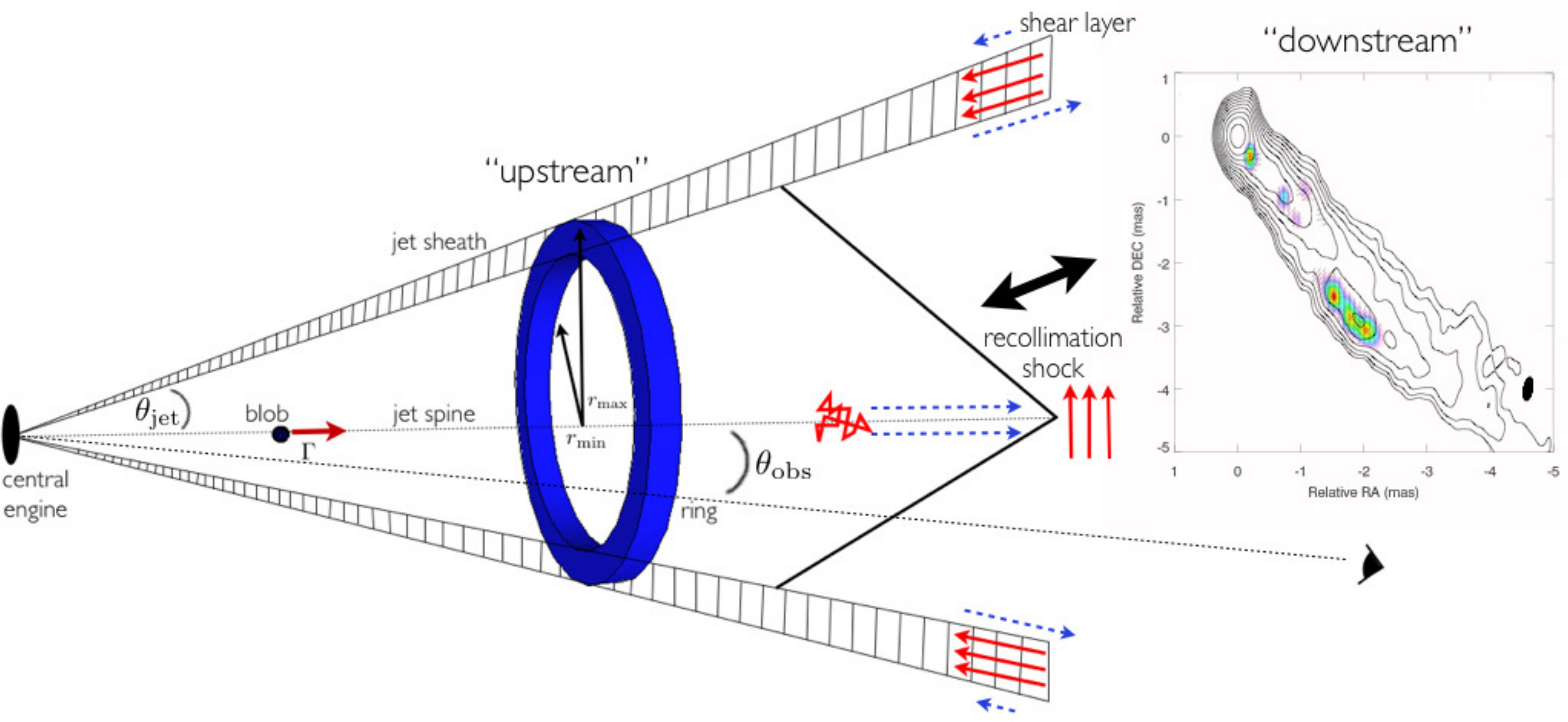}}
  \end{center}
  \caption{\label{fig1}A schematic of the relative locations along the jet of both the ring of shocked sheath plasma in our model and the location of the radio core/sheath detected farther ``downstream'' in the stacked radio images of 3C 273 (see Figures \ref{fig2} and \ref{fig3}).  This sketch is projected onto the plane of the sky.  We posit that the ring is located $\sim 4 ~ \rm{pc}$ from the central engine, while the radio cores in our stacked maps are located farther downstream from the central engine (at a scale of $\gtrsim 10 ~ \rm{pc}$).  We propose that the radio core in 3C 273 is associated with a recollimation shock that compresses initially tangled magnetic field along the spine of the jet and orders that field perpendicular to the jet axis (the red vectors just to the right of the recollimation shock).  The jet has an opening angle $\lesssim 2^{\circ}$ and the recollimation shock subtends an angle to the jet axis $\lesssim10^{\circ}$.  In contrast to the spine, velocity shear between the sheath and the ambient medium (blue vectors denote relative speed) aligns the magnetic field lines on the outer edges of the jet to be roughly parallel to the jet axis, resulting in the spine-sheath polarization signature we detect in our stacked map of 3C 273 (shown in the ``downstream'' portion of this Figure and in Figure \ref{fig3}).}
\end{figure*}

\cite{macdonald15} have developed the \textit{Ring of Fire} model to investigate the source of orphan $\gamma$-ray flares from within a blazar jet.  In this model, a segment (or ring) of a shocked non-relativistic sheath of plasma produces a concentrated region of synchrotron photons within the jet.  These photons are inverse-Compton scattered up to high energies by electrons contained within a blob of plasma propagating along the spine of the jet at superluminal speeds (see ``upstream'' portion of Figure \ref{fig1}).  The inverse-Compton scattering of the ring photons by the blob electrons creates an orphan $\gamma$-ray flare as the blob passes through the ring of shocked sheath.  This model is able to successfully reproduce a prominent orphan $\gamma$-ray flare observed within the blazar PKS 1510$-$089 (see \citealt{marscher10}).

In order to verify the existence of a sheath of plasma surrounding the spine of the jet in PKS 1510$-$089, \cite{macdonald15} implemented a method of radio map stacking (in Stokes I, Q, and U) in order to detect a tenuous polarimetric signal produced by the jet sheath, which is inherently less luminous than the highly Doppler beamed spine.  As discussed in \cite{wardle94}, if the jet sheath represents a ``shear layer'' between the relativistic jet and the environment into which the jet expands, one would expect the shear between the jet and the surrounding environment to ``stretch out'' the jet's initially helical (or tangled) magnetic field and align it parallel to the jet axis.  The observational characteristic of a shear layer would then be an increase in the observed fractional linear polarization towards the edges of the jet, with the orientation of that polarization being indicative of magnetic field that is aligned parallel to the jet axis.  The orientation of polarized radio emission is delineated by the electric vector position angle (EVPA).  The  EVPA angle ($\chi$) within a given pixel is computed based on the values of Stokes Q and U within that pixel: $\chi = \frac{1}{2} ~ \rm{arctan}( ~ U/Q ~ )$.  Neglecting the influence of relativistic aberration and Faraday rotation (\citealt{lyutikov05}), the EVPA should be aligned predominantly perpendicular to the projection of the magnetic field onto the plane of the sky.  Therefore, increased polarization towards the edges of the jet, with EVPA aligned roughly perpendicular to the jet axis, can be interpreted as a polarimetric footprint of a jet sheath.  This signal was detected in the stacked radio map of PKS 1510$-$089.

In this paper, we extend our modeling of orphan $\gamma$-ray flares and analysis of radio maps to a larger sample of blazars.  In particular, we search for orphan $\gamma$-ray flares in a sample of blazars monitored by the BU blazar group: www.bu.edu/blazars/VLBAproject.html.  We calculate $\gamma$-ray light curves using data provided by the Fermi Gamma-ray Space Telescope and construct optical light curves in R band using ground based telescopes around the world (for details see \citealt{williamson14}).  We define an orphan $\gamma$-ray flare as a $\gamma$-ray flare that does not have an optical counterpart.  

This paper is organized as follows: In \S2 we summarize the \textit{Ring of Fire} model and outline the model parameters (a detailed description of which can be found in \citealt{macdonald15}).  In \S3 we describe our method of radio map stacking.  In \S4 we present our stacked radio maps and orphan $\gamma$-ray flare models for our sample of blazars.  \S5 contains our summary and conclusions.  We adopt the following cosmological parameters: $H_{\rm o} = 70 ~ \rm km ~ \rm s^{-1} ~ \rm Mpc^{-1}$, $\Omega_{\rm m} = 0.3$, and $\Omega_{\Lambda} = 0.7$. 

\section{The Ring of Fire Model}

\begin{figure*}
  \setlength{\abovecaptionskip}{-6pt}
  \begin{center}
    \scalebox{0.84}{\includegraphics[width=2.0\columnwidth,clip]{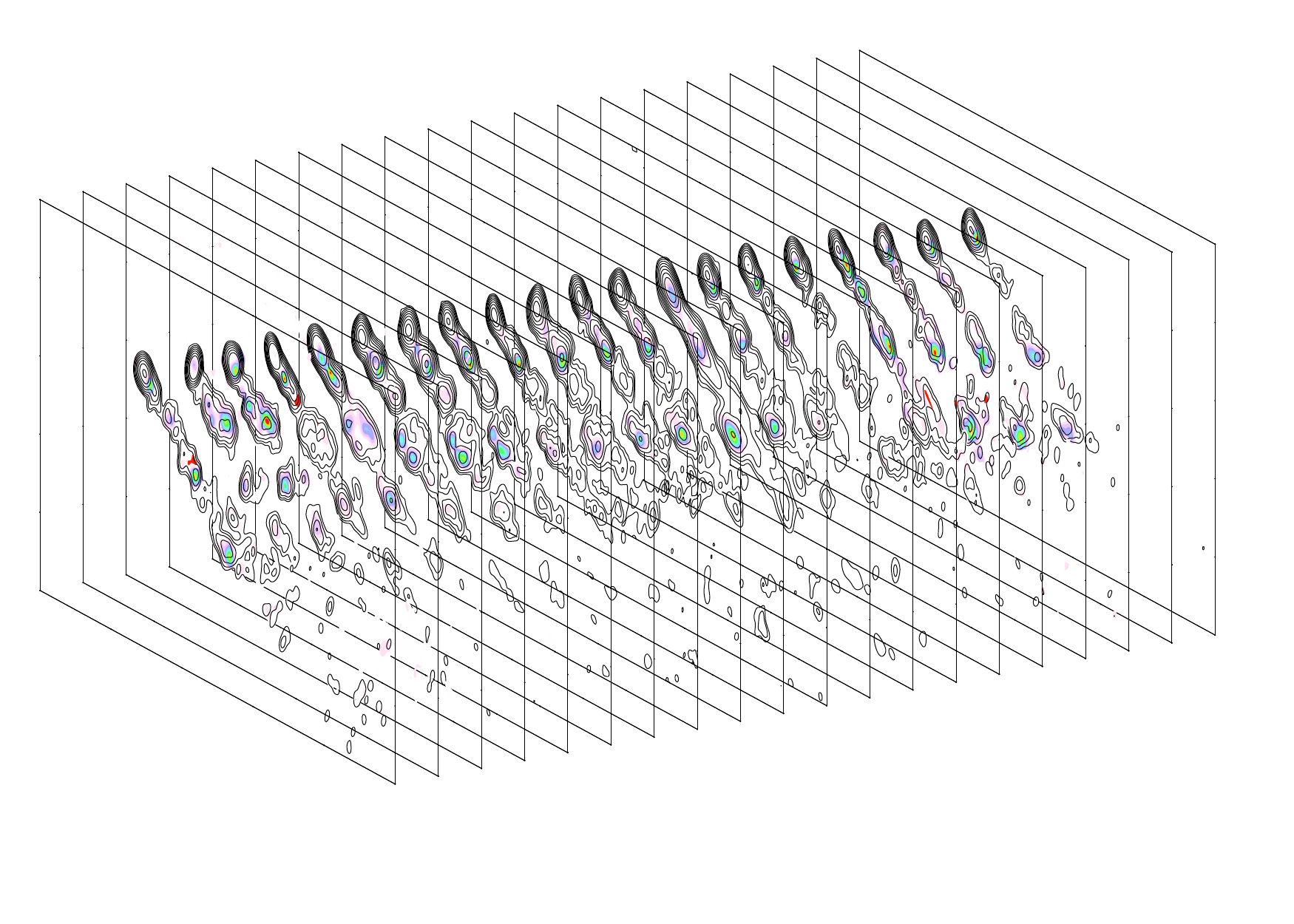}}
  \end{center}
  \vspace*{-1.7cm}
  \caption{\label{fig2}A schematic representation of the process of stacking radio maps for the blazar 3C 273.  For each of the blazars in our sample, we selected roughly twenty epochs of relatively quiescent jet activity.  The VLBA data were obtained from 2008 to 2016 as part of the VLBA-BU-BLAZAR program (see www.bu.edu/blazars/VLBAproject.html and \citet{jorstad13} for a general description of the observations and data analysis).  For each epoch we used Difmap's \emph{modelfit} task to locate the radio core.  The images were then aligned (based on the location of the radio core) and added together to create stacked I, Q, and U maps (shown in Figure \ref{fig3}).}
\end{figure*}

\begin{deluxetable}{lc}
\tablecolumns{2}
\tablewidth{1.0\columnwidth}
\tabletypesize{\normalsize}
\tablecaption{Ring of Fire Global Parameters \label{tab1}}
\startdata
\tableline
\noalign{\smallskip}
\tableline
\noalign{\smallskip}
Electron Power-Law Parameters & Value \\
\tableline
\noalign{\smallskip}
$\gamma_{\rm{min}}$ & $2.0 \times 10^{3}$ \\
\noalign{\smallskip}
$\gamma_{\rm{max}}$ & $1.0 \times 10^{4}$ \\
\noalign{\smallskip}
$s$ & 4.0 \\
\noalign{\smallskip}
\tableline
\noalign{\smallskip}
Blob Parameters & Value \\
\tableline
\noalign{\smallskip}
$r_{\rm{blob}}$ & 0.09 (pc)  \\
\noalign{\smallskip}
$B_{\rm{blob}}$ & 0.02 (G) \\
\noalign{\smallskip}
\tableline
\noalign{\smallskip}
Ring Parameters & Value \\
\tableline
\noalign{\smallskip}
$r_{\rm{min}}$ & 0.09 (pc) \\
\noalign{\smallskip}
$r_{\rm{max}}$ & 0.18 (pc) \\
\noalign{\smallskip}
$h_{\rm{o}}$ & 0.018 (pc) \\
\noalign{\smallskip}
$B_{\rm{ring}}$ & 0.12 (G) \\
\vspace*{-0.2cm}
\enddata
\end{deluxetable}

The \textit{Ring of Fire} model incorporates two blazar emission zones.  The two zones interact radiatively through inverse-Compton scattering.  The ring supplies seed photons that are up-scattered to $\gamma$-ray energies by a power-law distribution of electrons located within the blob.  The electron power-law energy distribution (both in the ring and in the blob) is set to $n_{e}(\gamma) \propto \gamma^{-s}$, over the energy range  $\gamma_{\rm min}$ to $\gamma_{\rm max}$, where the energy of an individual electron is $E = \gamma m_{e} c^{2}$.  The emission is computed in the co-moving frame of the blob, after which a number of Lorentz transformations are applied in order to generate the flux in the observer's frame.  In particular, we compute three different photon production rates accounting for synchrotron and synchrotron self-Compton photons produced internally to the blob and external Compton photons produced by the blob's passage through the photon field of the ring.  The distribution of electrons in the blob is evolved each time step by solving the Fokker-Planck equation (Kardashev 1962), thus accounting for radiative losses as the blob propagates through the ring.  We include an injection power ($P_{\rm inj}$) that parameterizes a physical mechanism at work within the blob (e.g., turbulence or diffusive shock acceleration), which continuously reenergizes the aging blob electron distribution.  The blob is assumed to be spherical with a radius $r_{\rm{blob}}$.  The ring of shocked sheath is parameterized by an inner ($r_{\rm{min}}$) and outer ($r_{\rm{max}}$) radius and a thickness ($h_{\rm{o}}$).  Uniform magnetic fields are set in the blob ($B_{\rm{blob}}$) and in the ring ($B_{\rm{ring}}$), both of which are assumed constant over the time it takes the blob to pass through the ring.  The blob begins its acceleration towards the ring at $z_{\rm{initial}}$ (along the jet axis) with an initial bulk Lorentz factor $\Gamma_{\rm{initial}}$ ($v_{\rm blob} \equiv c ~ \sqrt{ 1 - \Gamma^{-2} }$).  The blob ceases its acceleration at $z_{\rm{final}}$ after which it continues down the jet at a constant speed of $\Gamma_{\rm{final}}$.  The ring is situated at $z_{\rm{ring}} = 0 ~ \rm{pc}$ (the origin of our model), and although in reality the sheath (and our ring) will necessarily have some velocity parallel to the jet spine, for the simplicity of this calculation the ring is treated as static.  Finally, we include a fixed level of flux in the optical ($\rm{Baseline} ~ \rm{Flux}_{~ \rm{optical}}$) and $\gamma$-ray ($\rm{Baseline} ~ \rm{ Flux}_{~ \gamma-\rm{ray}}$) bands, to which the time variable emission produced by our model is added.  These static fluxes mirror the more slowly varying flux of the jet and are assumed to be constant over the course of our simulations.  The parameters $\Gamma_{\rm{initial}}$, $\Gamma_{\rm{final}}$, $z_{\rm{initial}}$, $z_{\rm{final}}$, $P_{\rm inj}$, the baseline fluxes, the angle of inclination of the jet to our line-of-sight ($\theta_{\rm{obs}}$), and the redshift ($Z$) are all source/flare specific and are listed in Table \ref{tab2} for each source considered.  The rest of the model parameters are fixed for all of the calculations presented below, and their specific values are listed in Table \ref{tab1}.  For a more in-depth description of the physics included in these calculations, see \cite{macdonald15}.  

\section{Radio Map Stacking}    

In order to detect the faint polarimetric signature of a jet sheath, we implement a method of radio map stacking (see, e.g., \citealt{fromm13}; \citealt{zamaninasab13}).  For each blazar in our sample, we stack roughly twenty uniformly weighted radio images obtained at a frequency of 43 GHz with the Very Long Baseline Array (VLBA) as part of the VLBA-BU-BLAZAR program (see www.bu.edu/blazars/VLBAproject.html). Each of the images make an equal contribution to the final stacked map. In particular, we selected epochs from relatively quiescent periods of jet activity (see Appendix A).  The aim of stacking the images is to ``smooth'' out transient features (like blobs propagating down the jet) and to ``amplify'' more static structures in the jet (like a jet sheath).  Each image is carefully aligned within the stack based on the location of the radio core at 43 GHz.  At the majority of our epochs, the radio core is the brightest feature in the jet; however, there are epochs where the flux from a blob downstream exceeds that of the radio core.  In order to avoid misidentification of the radio cores at the various epochs included in our stacked maps, we use the Difmap procedure \emph{modelfit} to identify the core among the features in each radio map.  We then use an IDL code to align and stack the images (see Figure \ref{fig2}) based on the location of the radio core in each image.  By stacking these images we are able to build up a radio signal from the periphery of each jet in our sample where the plasma sheath is hypothesized to exist.       

\begin{figure*}
  \setlength{\abovecaptionskip}{-6pt}
  \begin{center}
    \scalebox{1.0}{\includegraphics[width=2.0\columnwidth,clip]{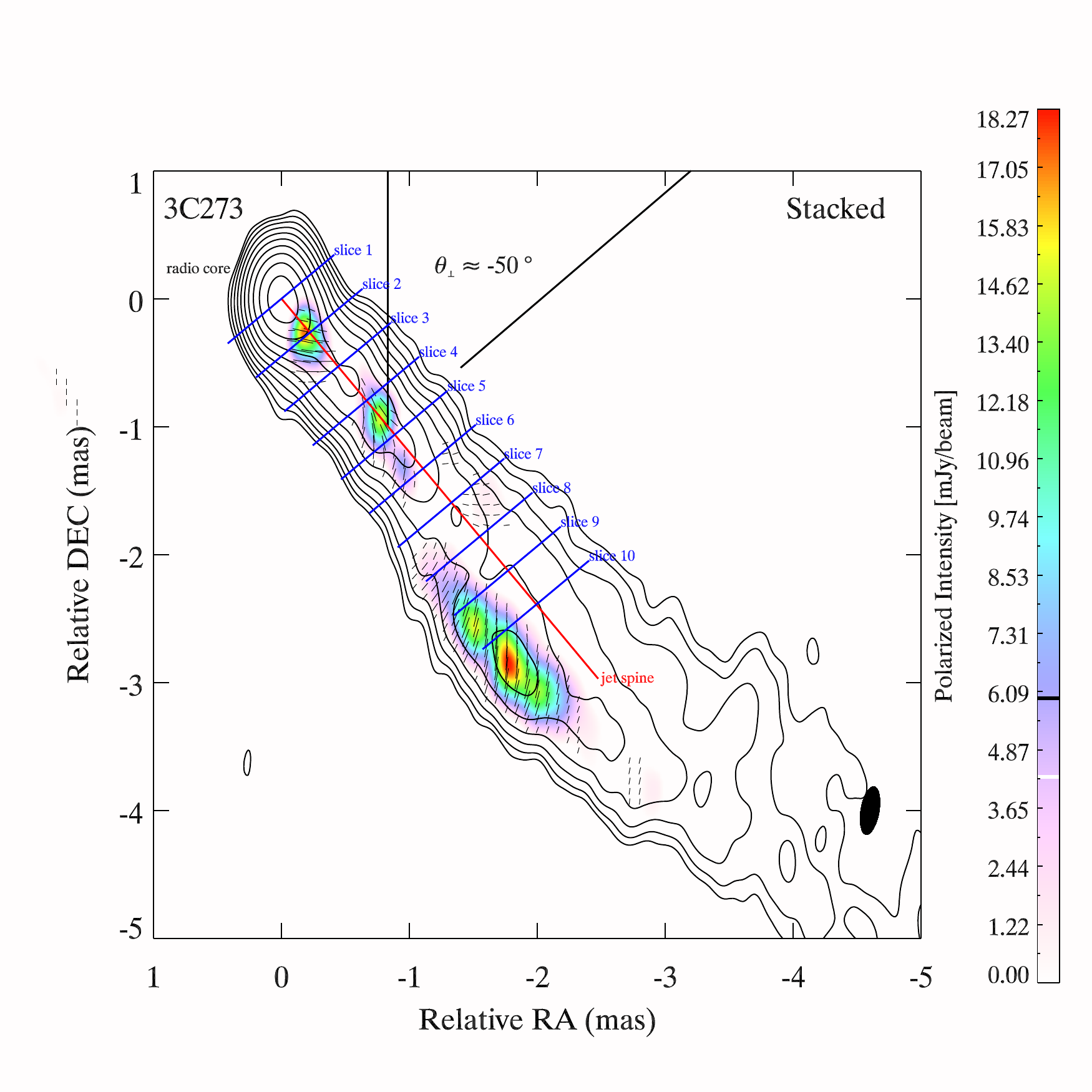}}
  \end{center}
  \caption{\label{fig3}A stacked map of 43 GHz images of 3C 273 spanning twenty epochs of observation from 2008 to 2016.  The black contours correspond to total intensity (contour levels are: 3.8, 7.5, 15.0, 30.0, 60.1, 120.2, 240.5, 480.9, 961.9, 1923.9 mJy beam$^{-1}$), whereas the underlying color scheme corresponds to polarized intensity (see color bar to the right for the flux levels), with the EVPAs denoted by black line segments.  The EVPAs indicate the orientation of linear polarization as projected onto the plane of the sky.  All images have been convolved with a Gaussian beam, shown in the bottom right corner of the stacked map.  A jet spine is plotted in red based on the transverse symmetry of the jet, across which ten transverse slices through the data are taken (shown in blue).  The profiles of the emission parameters along the sixth slice are shown in Figure \ref{fig5}.  The two black lines highlight the angle ($\theta_{\perp}$) between $0^{\circ}$ (i.e. North) and $90^{\circ}$ relative to the jet spine (in red).  This transverse orientation relative to the jet spine is plotted as a horizontal black line in Figure \ref{fig5} to highlight EVPAs that are oriented roughly perpendicular to the jet axis.}
\end{figure*}

\newpage

\section{Results} 

\subsection{\rm{3C 273}}

For an initial test of our stacking procedure's ability to detect jet sheaths, we selected the blazar 3C 273.  At a redshift of $Z = 0.158$ (\citealt{lavaux11}), 3C 273 is one of the closest examples of a blazar, and as such is an ideal candidate to search for a jet sheath.  As outlined in \S3, we selected twenty relatively quiescent epochs of jet activity to create the stacked radio map shown in Figure \ref{fig3}.  As predicted by the spine-sheath model, the linear polarization increases toward the edges of the jet downstream of the radio core. The EVPAs downstream are predominantly inclined to the jet spine (shown as a red line delineating the historical trajectories taken by blobs ejected from the radio core).  A series of slices was taken through our stacked map to highlight the profiles of the Stokes parameters transverse to the jet axis.      

\begin{figure}
\epsscale{1.03}
\plotone{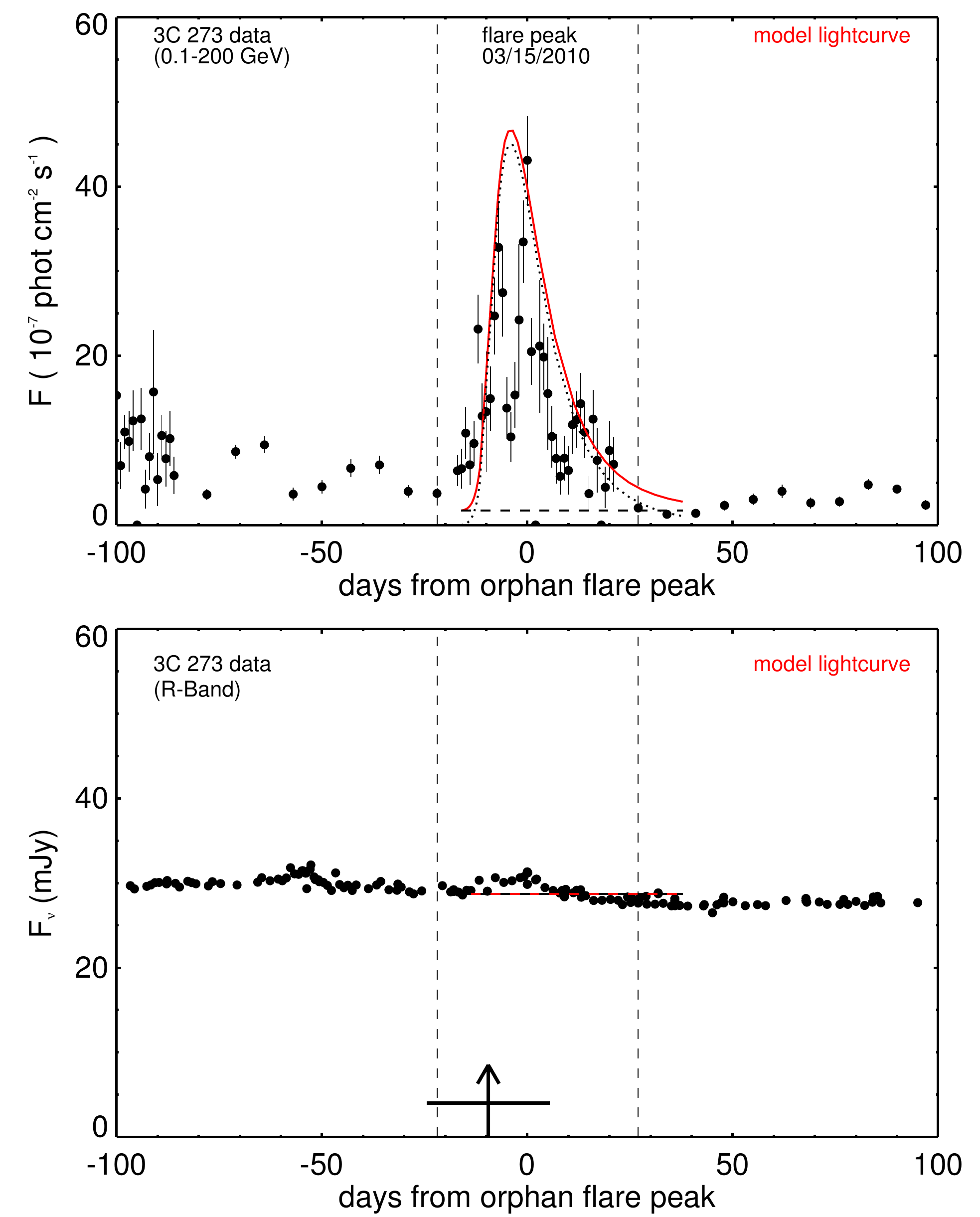}
\caption{\label{fig4}Light curves of 3C 273 (black circles) in the $\gamma$-rays (upper panel) and optical (lower panel).  The model light curves are overlaid in red and consist of the superposition of the model baseline level of flux in each band (the horizontal dashed black line) and the variable emission produced by the model (the dotted black line).  The vertical arrow in the lower panel marks the time when a superluminal knot passed through the 43 GHz core of 3C 273, with the horizontal bar representing the uncertainty in this time.}
\end{figure}

We detect an orphan $\gamma$-ray flare in the Fermi LAT light curve during March 2010.  The corresponding light curves in $\gamma$-rays and in the optical band are shown in Figure \ref{fig4}.  The VLBA images also exhibit the ejection of a blob from the radio core of 3C 273 during the onset of this orphan flare on March 6th 2010 $\pm$ 25 days (\citealt{jorstad17}), with a bulk Lorentz factor of $\Gamma \sim 6$.  The time of the blob ejection is demarcated by a vertical arrow in the lower panel of Figure \ref{fig4}.  Table \ref{tab2} lists the \textit{Ring of Fire} model parameters used to obtain a fit to this orphan $\gamma$-ray flare.  The synthetic $\gamma$-ray and optical light curves produced by the \textit{Ring of Fire} model are overlaid in red.  The general strategy used to fit the orphan flares presented within this paper was to start with the parameters used in \cite{macdonald15} to model the orphan $\gamma$-ray flare from PKS 1510$-$089.  As discussed in \cite{macdonald15}, the shape of the orphan flare profile produced by the \textit{Ring of Fire} model is very sensitive to the nature of the blob's acceleration down the jet spine.  With this in mind, the ring parameters were held constant and only the nature of the blob's acceleration was modified to obtain the above fit to the data.  Due to the larger inclination angle of the jet in 3C 273 to our line of sight ($\theta_{\rm{obs}} = 6.1^{\circ}$ in contrast to PKS 1510$-$089 where $\theta_{\rm{obs}} = 1.4^{\circ}$; \citealt{jorstad05}), the power ($P_{\rm{inj}}$) injected into the blob was also increased in order to match the observed flux levels.  The bulk Lorentz factor used in our model ($\Gamma \sim 25$) is much larger than the apparent speed of the knot above.  The disparity in these values highlights either (i) a shortcoming in our model or (ii) some form of (de)acceleration experienced by the blob upstream of the core (see, e.g., \citealt{georganopoulos03}).       

\begin{figure}
  \setlength{\abovecaptionskip}{-6pt}
  \begin{center}
  \hspace*{-1.4cm}
    \scalebox{0.81}{\includegraphics[width=1.4\columnwidth,clip]{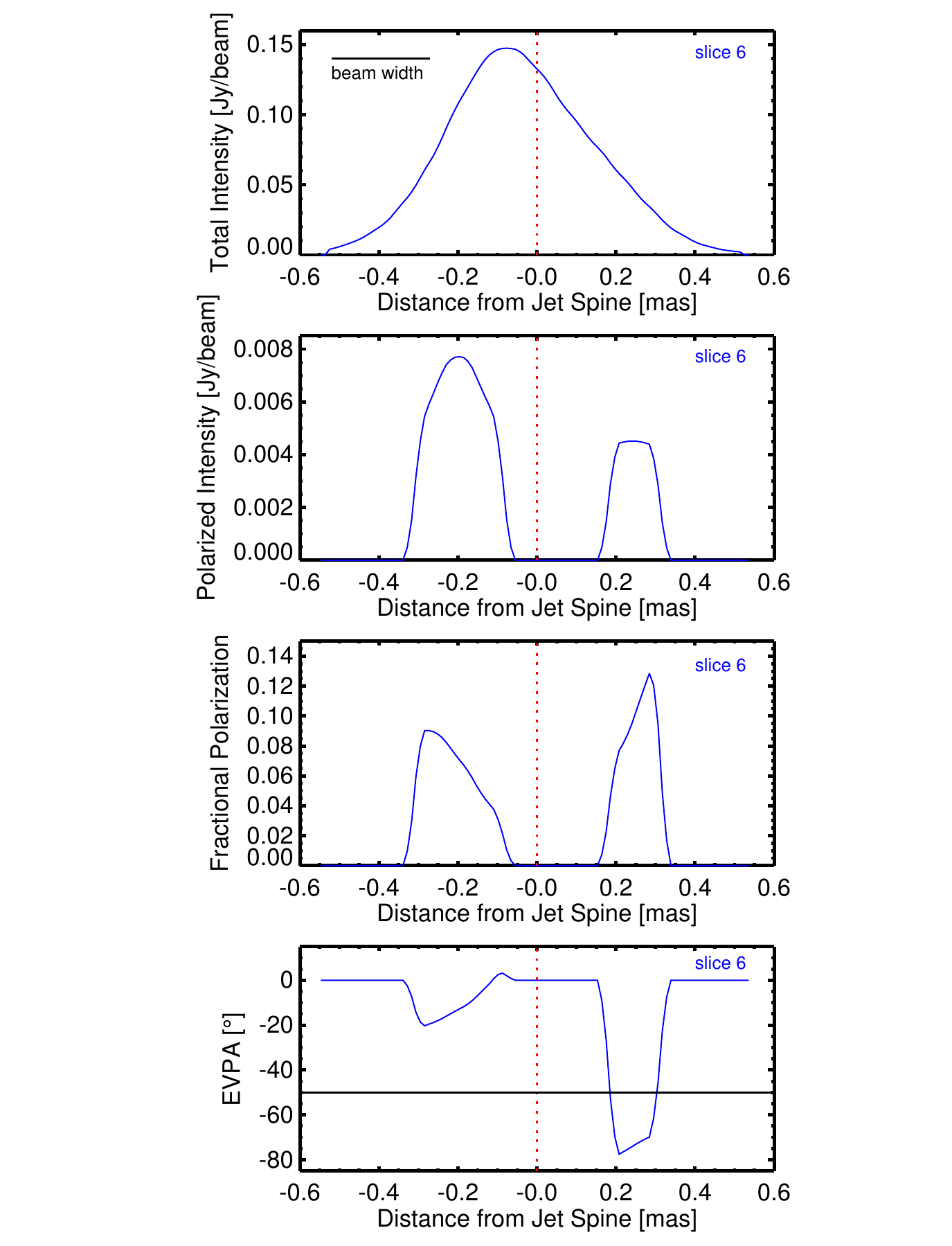}}
  \end{center}
  \caption{\label{fig5}Variations of the emission parameters of 3C 273 transverse to the jet axis (dashed red line) for slice 6 (shown in Figure \ref{fig3}) from top to bottom: total intensity, polarized intensity, fractional polarization, and EVPA.  The fractional polarization increasing toward the edges of the jet is a predicted polarimetric signature of a jet sheath (as discussed in \S1), with the EVPA transverse to the axis.  The black line in the lower panel delineates EVPAs that are exactly $\perp$ to the jet spine. The beam width along this transverse slice is shown in the upper panel to the left.}
\end{figure}
  \vspace*{-0.02cm}
Figure \ref{fig5} shows the profiles of the Stokes parameters (I, Q, and U) transverse to the jet axis along the sixth slice through our stacked map (Figure \ref{fig3}).  The upper middle panel highlights the double peaked profile in the linearly polarized intensity ($\sqrt{ Q^{2} + U^{2} }$).  The lower middle panel shows how the fractional polarization ($\sqrt{ Q^{2} + U^{2} }/I$) increases towards the edges of the jet, in agreement with the spine-sheath model. There are rotation measure (RM) gradients across the jet of 3C 273 (see \citealt{asada02} and \citealt{zavala05}), however, that indicate that this increase in fractional polarization may also be indicative of a large-scale helical magnetic field (see \S \ref{stuff} for further discussion). We also point out that there is a polarized feature just downstream of the radio core in our stacked map (Figure \ref{fig3}, slice 2) with EVPAs aligned parallel to the jet spine.  This is a signature expected for a standing shock just downstream of the radio core that compresses and aligns the magnetic field (\citealt{cawthorne90}).  

\begin{figure*}
  \setlength{\abovecaptionskip}{-6pt}
  \begin{center}
    \scalebox{0.89}{\includegraphics[width=2.0\columnwidth,clip]{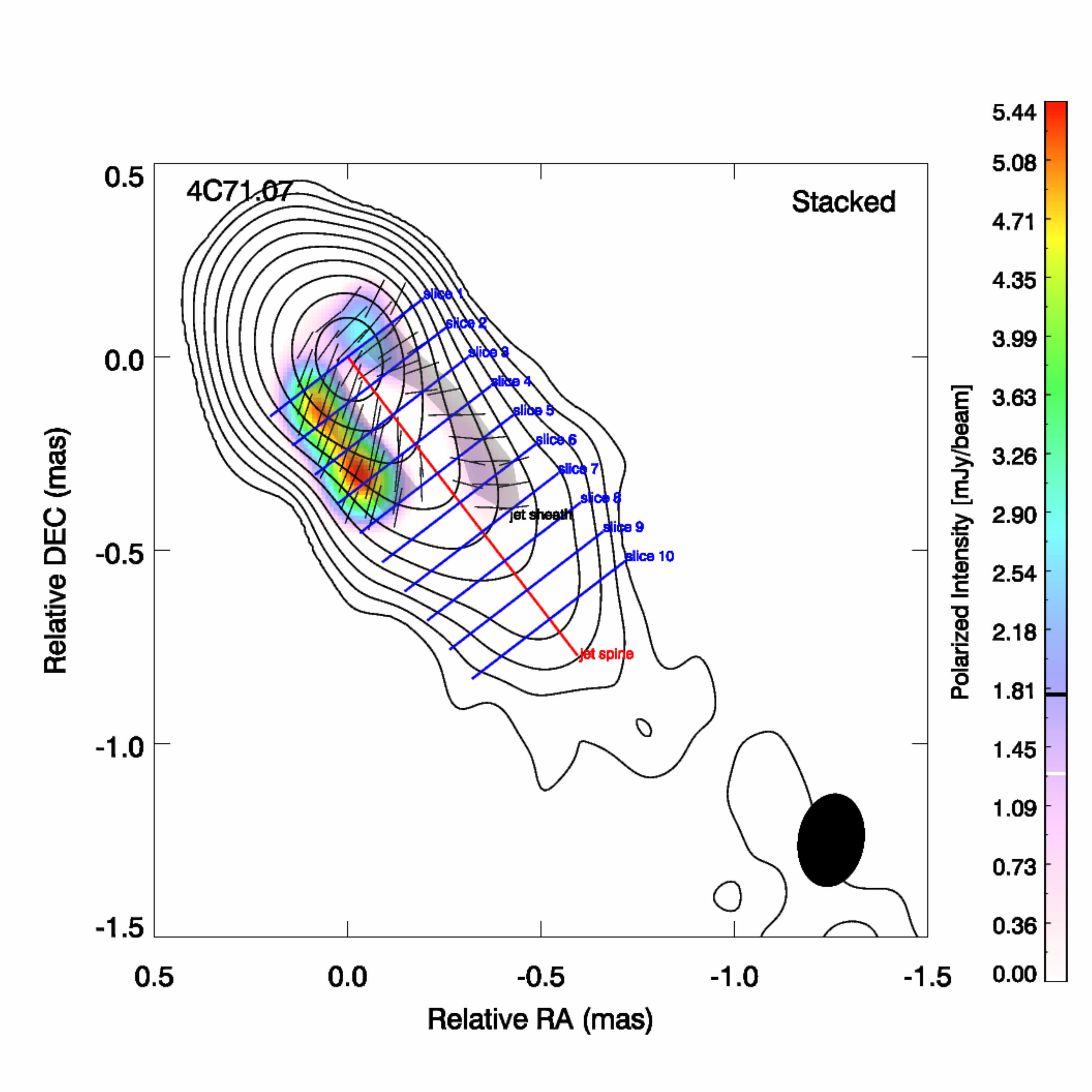}}
  \end{center}
  \caption{\label{fig6}A stacked map of 43 GHz images of 4C 71$.$07 spanning twenty epochs of observation from 2008 to 2016.  As in Figure \ref{fig3}, the black contours correspond to total intensity (contour levels are: 0.9, 1.9, 3.9, 7.9, 15.8, 31.7, 63.4, 126.9, 253.9, 507.8 mJy beam$^{-1}$), whereas the underlying color scheme corresponds to polarized intensity (see color bar to the right for the flux levels), with the EVPAs denoted by black line segments.  The EVPAs indicate the orientation of linear polarization as projected onto the plane of the sky.   All images have been convolved with a Gaussian beam, shown in the bottom right corner of the stacked map.  A jet spine is plotted in red based on the transverse symmetry of the jet, across which ten transverse slices through the data are taken (shown in blue).  The profiles of the emission parameters along the first slice are shown in Figure \ref{fig8}.  The shaded gray regions in the above map highlight the nominal location of the jet sheath of 4C 71$.$07 as determined by the procedure illustrated in Figure \ref{fig9}.}
\end{figure*}

\newpage

\subsection{\rm{4C 71$.$07}}

Figure \ref{fig6} presents a stacked radio map of the blazar 4C 71$.$07 (0836$+$710).  A very prominent and continuous polarimetric signature of a jet sheath is evident down the edges of the jet in this stacked radio image.  Given the continuous nature of the jet sheath on the image, we can estimate the bolometric luminosity of the sheath.  This is accomplished by first using the length of the polarized intensity profile along each of the slices to trace out the shaded gray regions shown in Figure \ref{fig9} (see also Figure \ref{fig6}).  The size of the sheath on each side of the slice is arbitrarily set by the location where the polarized flux falls below 0.85 times the peak value.  An estimate of the bolometric luminosity ($L_{\rm bol}$) of the sheath is determined by summing up all of the deconvolved `sheath' flux (see Figures \ref{fig8} \& \ref{fig9}) contained within the gray shaded regions of our stacked map.  We convert this extrapolated sheath flux at 43 GHz into a spectral luminosity and assume a power-law ($L_{\nu} \propto \nu^{-\alpha}$), which we then integrate over a range of frequencies that encompass the sheath's radiation to obtain:
\begin{equation}
L_{\rm bol} = \int_{\nu_{\rm{min}}}^{\nu_{\rm{max}}} L_{\nu} ~ \! \mathrm{d} \nu \sim 2 \times 10^{43} ~ \rm erg ~ \rm s^{-1},
\end{equation}
where we have adopted a spectral index of $\alpha \sim 1.0$ and limits of integration of $\nu_{\rm min} = 10^{9} ~ \rm Hz$ and $\nu_{\rm max} = 5 \times 10^{13} ~ \rm Hz$. Our estimate indicates that the jet sheath can potentially provide a substantial source of seed photons even at parsec scales along the jet. It should be pointed out, however, that the beam width is $\sim 0.2 ~ \rm{mas}$ along our transverse slices and that the sheath region we detect is within $\sim 0.05-0.15 ~ \rm{mas}$ of the jet spine. Convolution with the beam, therefore, adds a significant contribution of flux from the spine to our nominal sheath region. As illustrated in Figures \ref{fig8} \& \ref{fig9} (solid red lines) we attempt to remove this contaminating spine flux from our estimate. The bolometric luminosity given in Equation 1, however, should only be regarded as a rough estimate of the actual sheath luminosity. 

An estimate of the inner ($r_{\rm min}$) and outer ($r_{\rm max}$) radii of the sheath was made by averaging these parameters relative to the jet spine for each slice through the stacked map.  We find that $r_{\rm min} \sim 0.27 ~ \rm pc$ and $r_{\rm max} \sim 0.34 ~ \rm pc$.  These values are larger than the inner and outer radii used in our model of the ring (see Table \ref{tab1}) that we postulate to exist upstream of the radio core.

\begin{figure}
\epsscale{0.97}
\plotone{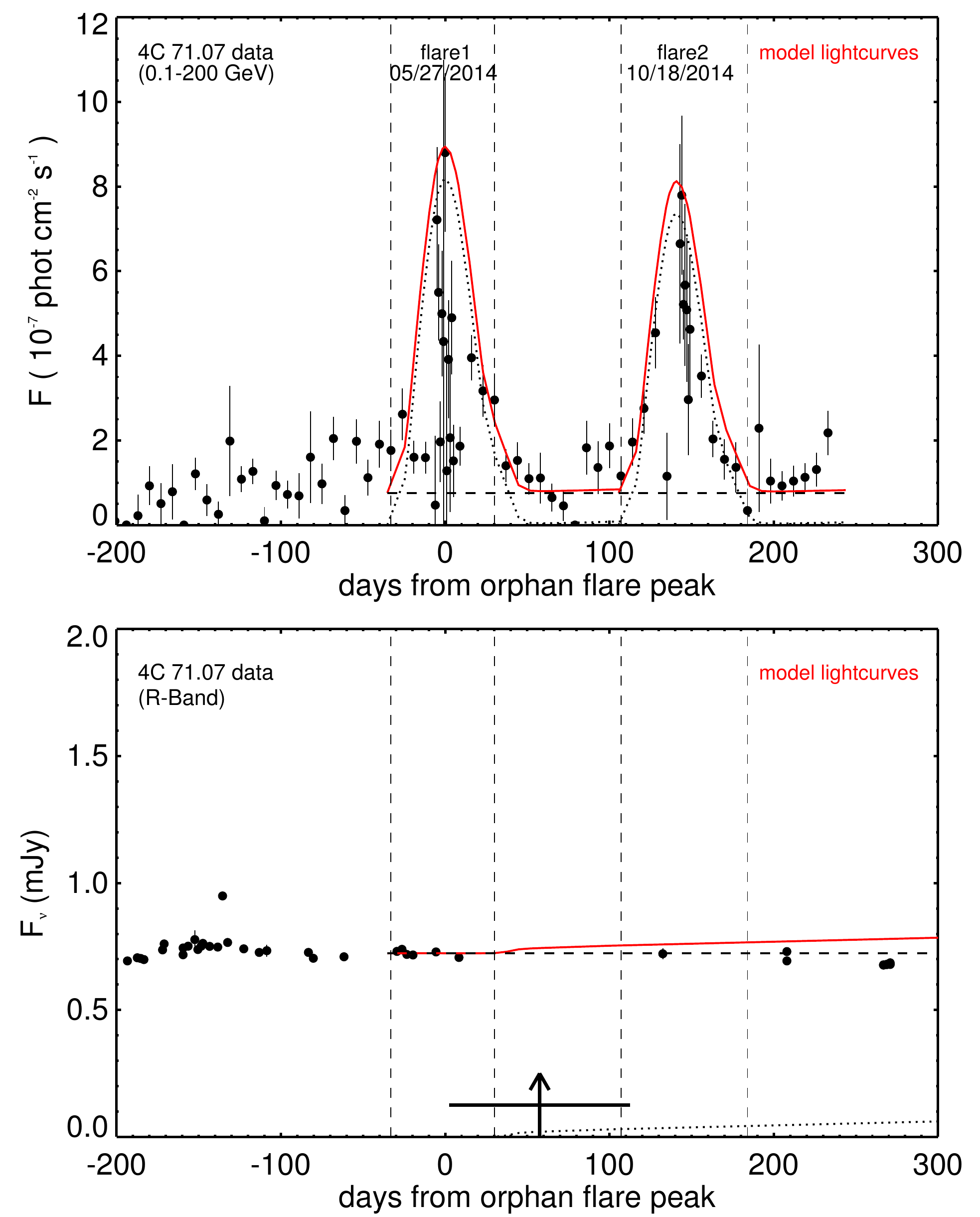}
\caption{\label{fig7}Light curves of 4C 71$.$07 (black circles) in the $\gamma$-rays (upper panel) and optical (lower panel).  The model light curves are overlaid in red and consist of the superposition of the model baseline level of flux in each band (the horizontal dashed black line) and the variable emission produced by the model (the dotted black line).  The vertical arrow in the lower panel marks the time when a superluminal knot passed through the 43 GHz core of 4C 71$.$07, with the horizontal bar representing the uncertainty in this time.}
\end{figure}  

We detect two orphan $\gamma$-ray flares in the Fermi LAT light curve of 4C 71$.$07 during 2014 (Figure \ref{fig7}), although the optical observations are sparse.  The first orphan flare occurred in May followed by a second in October.  Table \ref{tab2} lists the \textit{Ring of Fire} model parameters used to obtain a fit to this pair of orphan flares.  For this particular case, two consecutive blobs were launched through the ring of shocked sheath.  The nature of the blob acceleration down the jet spine was identical for the two blobs, with the only difference being a slightly smaller injection power used in the passage of the second blob through the ring.  The VLBA monitoring indicates the ejection of a blob from the 43 GHz core of 4C 71$.$07 on July 24th 2014 $\pm$ 55 days (Jorstad - private communication) with a bulk Lorentz factor of $\Gamma \sim 32$.  The bulk Lorentz factor in our model ($\Gamma \sim 10$) is too low compared to this VLBI speed.  Again, as in the case of 3C 273, this discrepancy potentially highlights some form of acceleration occurring within the jet (see, e.g., \citealt{marscher08}, \citealt{homan15}).  

\begin{figure}
  \setlength{\abovecaptionskip}{-6pt}
  \begin{center}
  \hspace*{-1.35cm}
    \scalebox{0.84}{\includegraphics[width=1.25\columnwidth,clip]{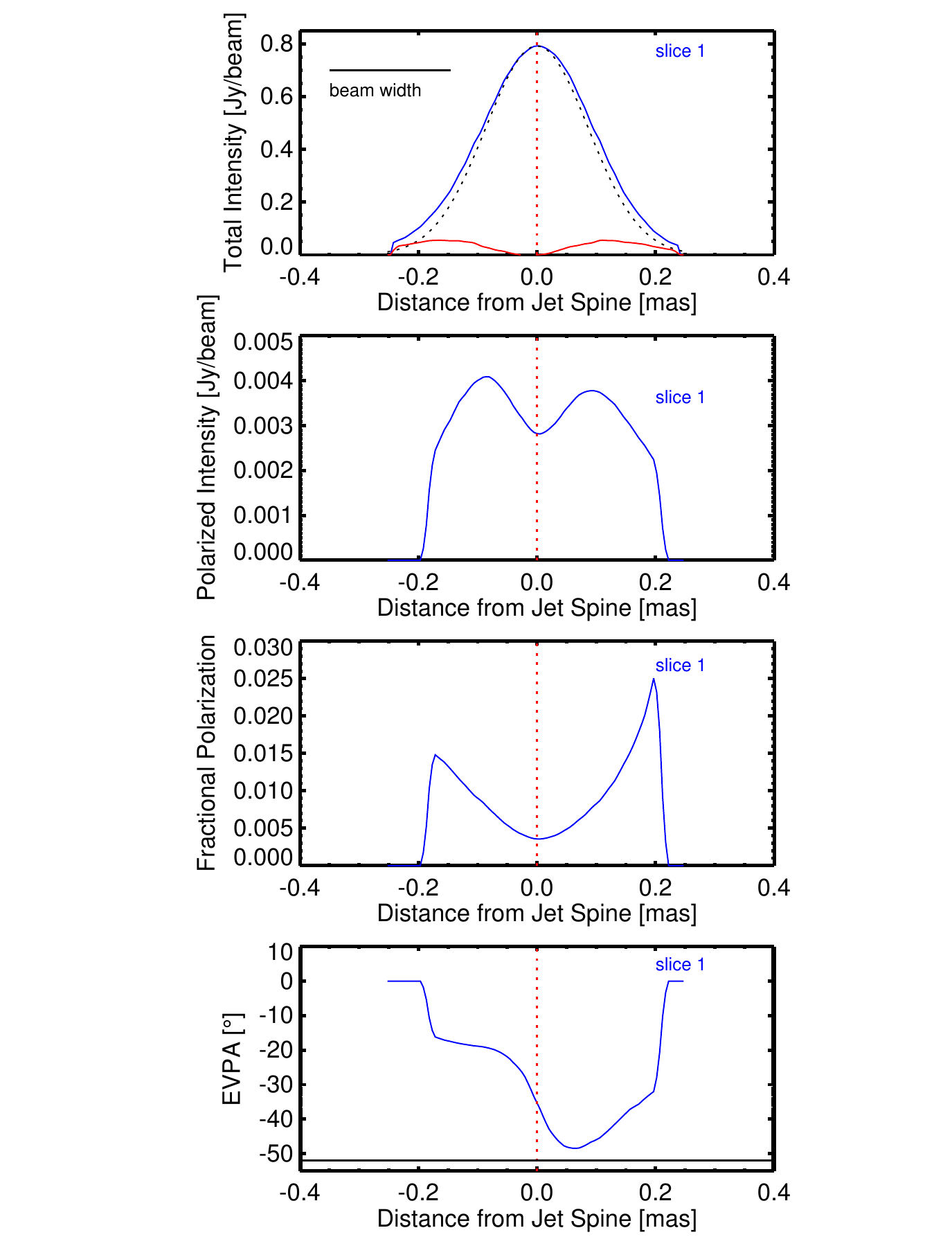}}
  \end{center}
  \vspace*{-0.2cm}
  \caption{\label{fig8}Variations of the emission parameters (solid blue lines) transverse to the jet axis (dashed red line) of 4C 71$.$07 for slice 1 (shown in Figure \ref{fig6}) from top to bottom: total intensity, polarized intensity, fractional polarization, and EVPA.  The fractional polarization increasing toward the edges of the jet is a predicted polarimetric signature of a jet sheath (as discussed in \S1), with the EVPA transverse to the axis.  The black line in the lower panel delineates EVPAs that are exactly $\perp$ to the jet spine. The beam width along this transverse slice is shown in the upper panel to the left. The beam profile (dashed black line) is deconvolved from the total intensity profile to estimate a `sheath' total intensity profile (solid red line).}
\end{figure}

\begin{figure}
  \setlength{\abovecaptionskip}{-6pt}
  \begin{center}
  \hspace*{0.36cm}
    \scalebox{0.93}{\includegraphics[width=0.83\columnwidth,clip]{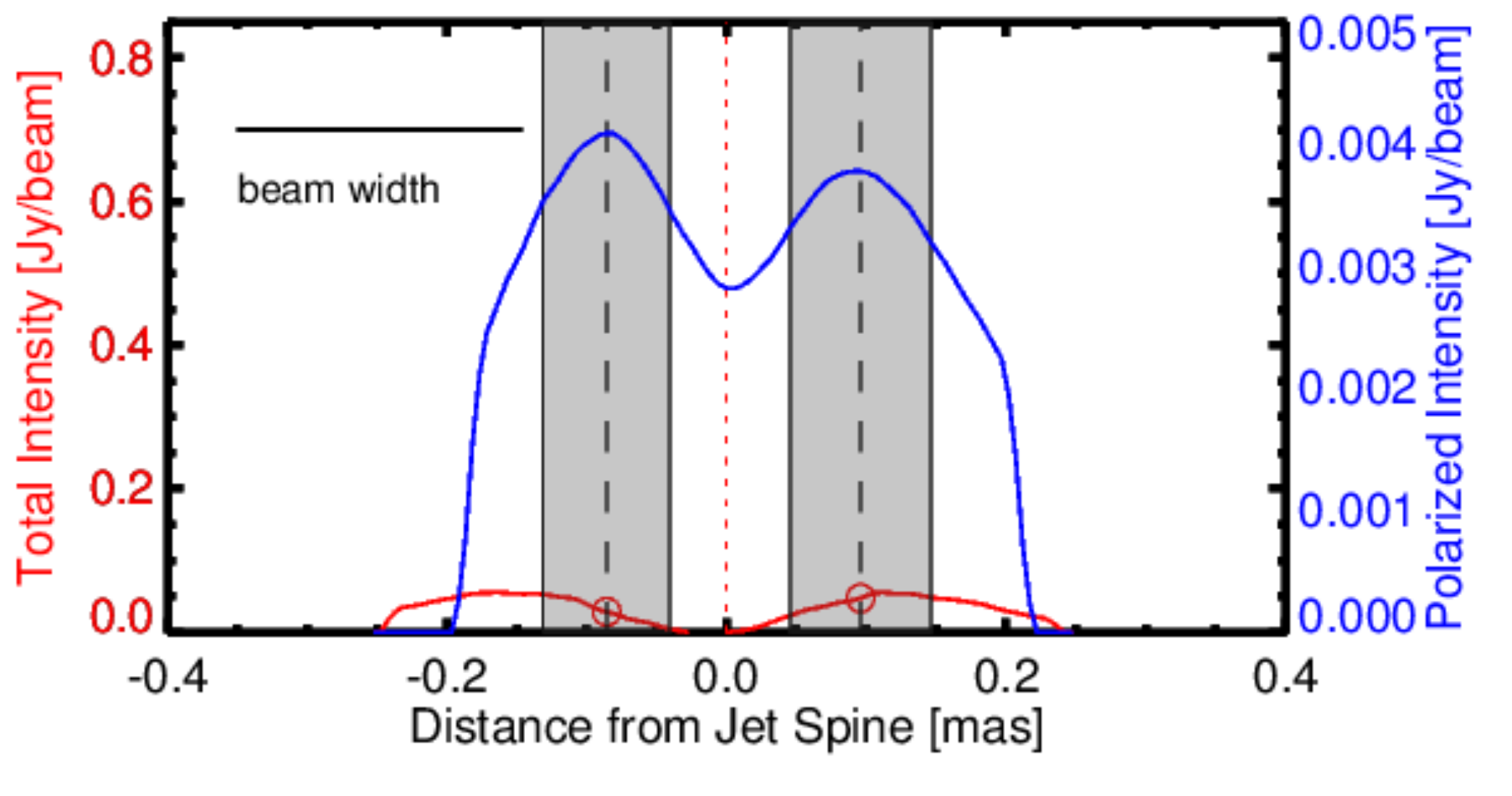}}
  \end{center}
  \vspace*{-0.3cm}
  \caption{\label{fig9}The polarized intensity profile of 4C 71$.$07 along slice 1 (solid blue line corresponding to the right-hand axis) is overlaid upon the deconvolved `sheath' total intensity profile for slice 1 (solid red line corresponding to the left-hand axis).  We use this double-peaked profile to highlight the likely location of the sheath within each slice (shaded gray regions).  The sheath's contribution to the jet's total intensity profile is estimated by the values of total intensity (red circles) that are cospatial with the peaks of the polarized intensity profile (demarcated by dashed vertical lines).  This procedure is then repeated for each slice, thus tracing out the sheath (shaded gray regions) shown in Figure \ref{fig6}.}
\end{figure}  

\begin{figure*}
  \setlength{\abovecaptionskip}{-6pt}
  \begin{center}
    \scalebox{1.0}{\includegraphics[width=2.0\columnwidth,clip]{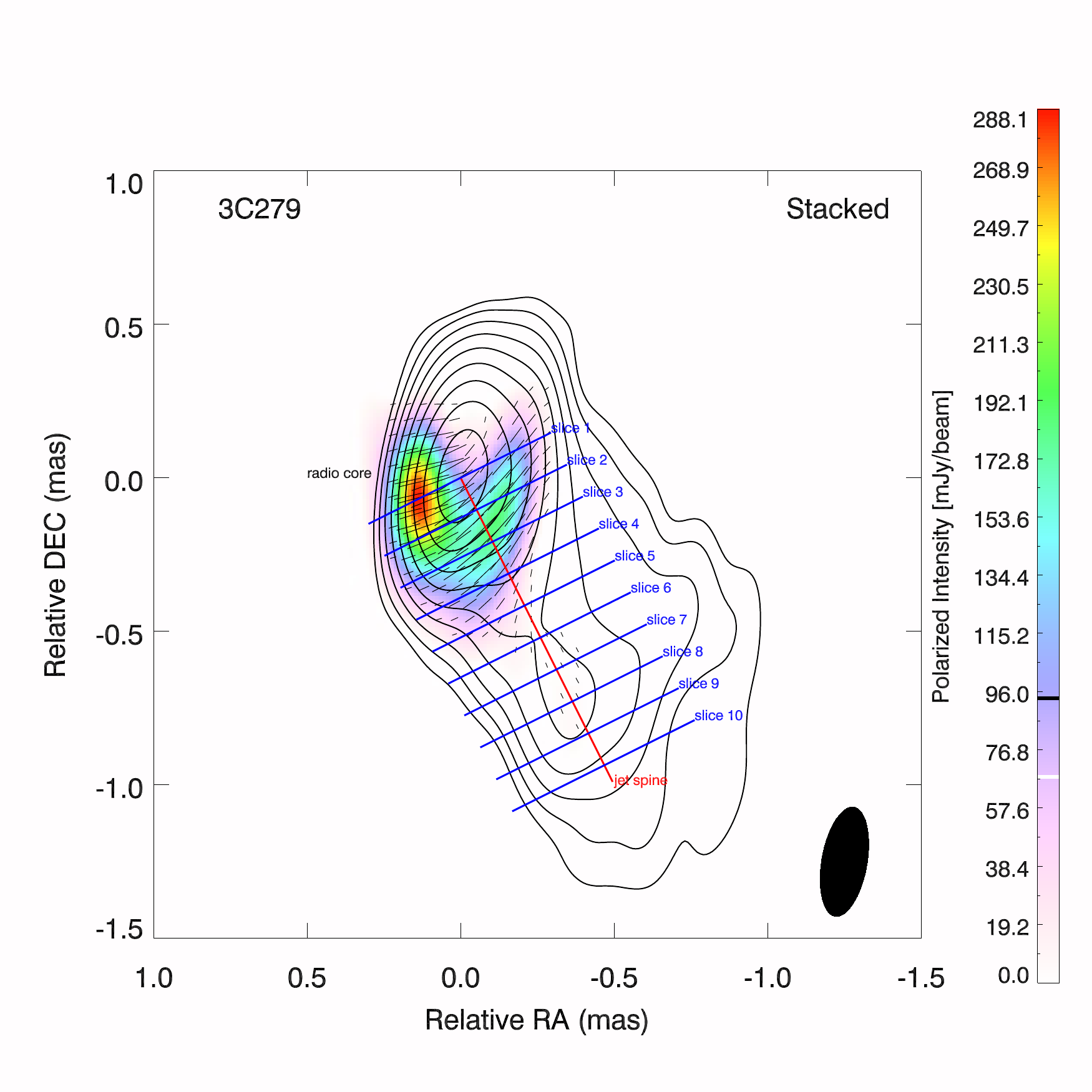}}
  \end{center}
  \caption{\label{fig10}A stacked map of 43 GHz images of 3C 279 spanning ten epochs of observation from 2008 to 2016.  The black contours correspond to total intensity (contour levels are: 20.5, 41.0, 82.0, 164.0, 328.0, 656.1, 1312.2, 2624.4, 5248.9 mJy beam$^{-1}$), whereas the underlying color scheme corresponds to polarized intensity (see color bar to the right for the flux levels), with the EVPAs denoted by black line segments.  The EVPAs indicate the orientation of linear polarization as projected onto the plane of the sky.   All images have been convolved with a Gaussian beam, shown in the bottom right corner of the stacked map.  A jet spine is plotted in red based on the transverse symmetry of the jet, across which ten transverse slices through the data are taken (shown in blue).  The profiles of the emission parameters along the first slice are shown in Figure \ref{fig12}.}
\end{figure*}

\newpage

\subsection{\rm{3C 279}}

In Figure \ref{fig10} we present a stacked radio map of the blazar 3C 279.  In contrast to the stacked radio maps of 3C 273 (Figure \ref{fig3}) and 4C 71$.$07 (Figure \ref{fig6}), our map of 3C 279 shows a far more complex polarimetric pattern down the length of the jet.  Initially, the EVPAs are aligned perpendicular to the jet axis in the vicinity of the radio core.  The polarization then bifurcates across the width of the jet, with the EVPAs remaining inclined predominantly perpendicular to the jet axis (especially on the edges of the jet).  The linearly polarized intensity, however, peaks to one side of the jet (see the upper middle panel of Figure \ref{fig12}, in which the ``double-peaked'' linearly polarized intensity profile is not as symmetric as those shown in Figures \ref{fig5} \& \ref{fig8}).  Farther downstream the EVPAs align with the jet axis and the polarization becomes confined to the jet spine. 

\begin{figure}
\epsscale{1.17}
\plotone{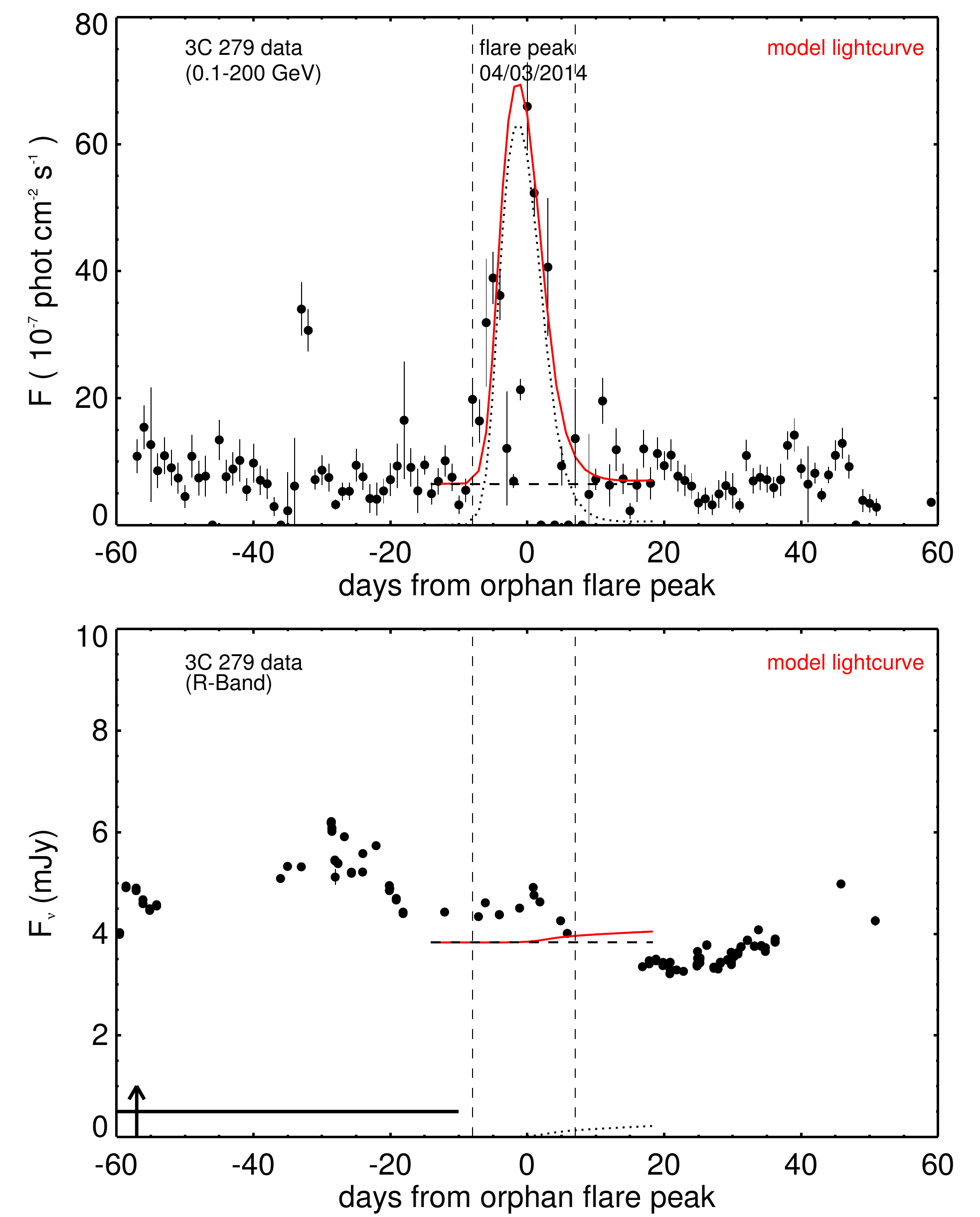}
\caption{\label{fig11}Light curves of 3C 279 (black circles) in the $\gamma$-rays (upper panel) and optical (lower panel).  The model light curves are overlaid in red and consist of the superposition of the model baseline level of flux in each band (the horizontal dashed black line) and the variable emission produced by the model (the dotted black line).  The vertical arrow in the lower panel marks the time when a superluminal knot passed through the 43 GHz core of 3C 279, with the horizontal bar representing the uncertainty in this time.}
\end{figure}

An orphan $\gamma$-ray flare with no similar optical counterpart is seen in the Fermi LAT light curve of 3C 279 during April 2014.  The corresponding $\gamma$-ray and optical light curves are shown in Figure \ref{fig11}.  Table \ref{tab2} lists the \textit{Ring of Fire} model parameters used to obtain a fit to this orphan $\gamma$-ray flare.  Again, the synthetic $\gamma$-ray and optical light curves produced by the \textit{Ring of Fire} model are overlaid in red.  There are several statistically significant data points in the $\gamma$-ray flare profile that are not reproduced by the \textit{Ring of Fire} model.  These three data points occur shortly after the onset of the orphan flare and result in a ``kink'' in the flare profile.  This kink could represent some catastrophic loss of electrons within the blob, perhaps as the relativistic electron injection mechanism at work within the blob (e.g., turbulence or magnetic reconnection) momentarily ceases.  This type of short-term discontinuity in the flare profile is not accounted for in the \textit{Ring of Fire} model, in which the internal conditions within the blob remain fixed over the course of the blob's passage through the ring of shocked sheath.  

\begin{figure}
  \setlength{\abovecaptionskip}{-6pt}
  \begin{center}
  \hspace*{-2.2cm}
    \scalebox{1.0}{\includegraphics[width=1.4\columnwidth,clip]{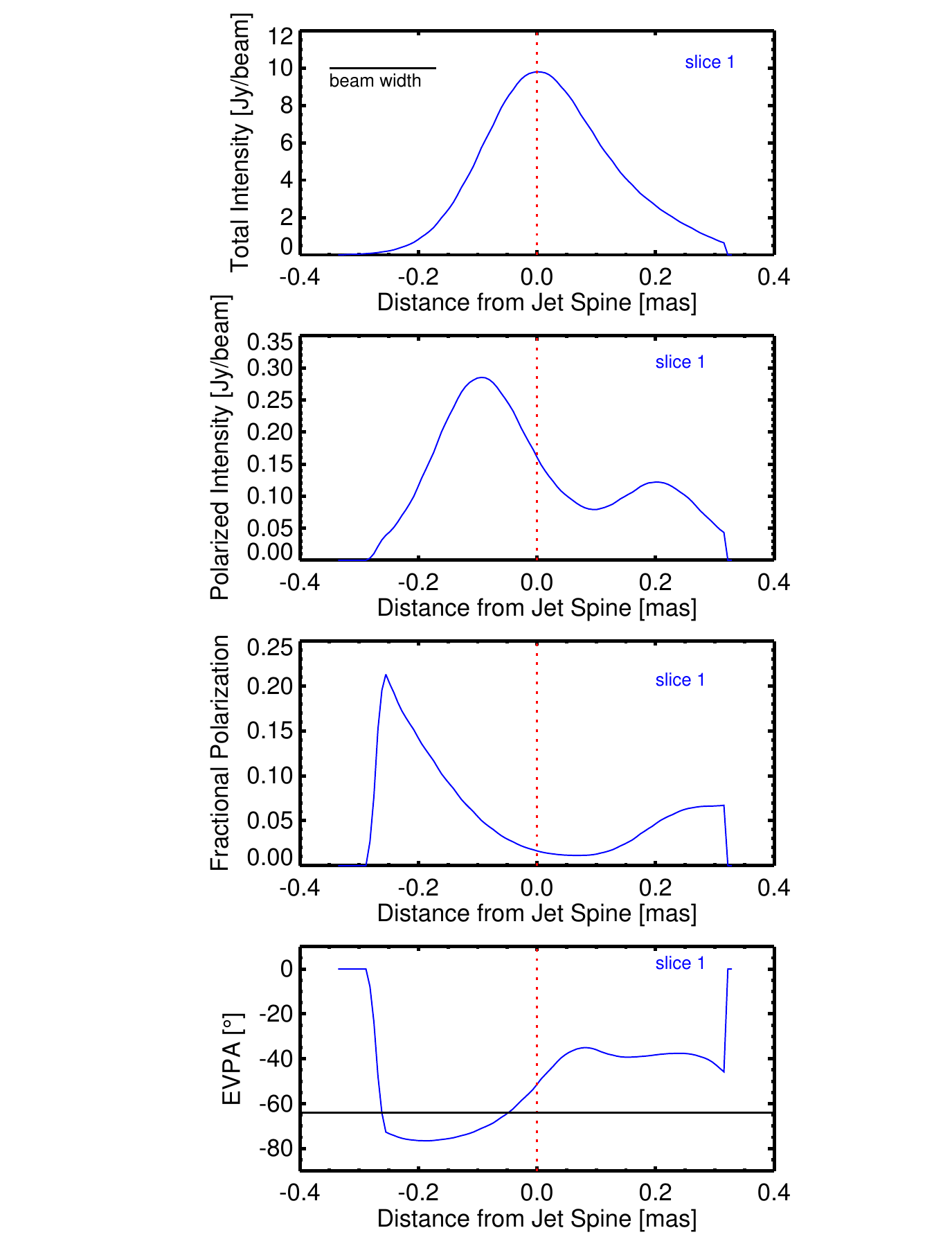}}
  \end{center}
  \caption{\label{fig12}Variations of the emission parameters of 3C 279 transverse to the jet axis (dashed red line) for slice 1 (shown in Figure \ref{fig10}) from top to bottom: total intensity, polarized intensity, fractional polarization, and EVPA.  The fractional polarization increasing toward the edges of the jet is a predicted polarimetric signature of a jet sheath (as discussed in \S1), with the EVPA transverse to the axis.  The black line in the lower panel delineates EVPAs that are exactly $\perp$ to the jet spine. The beam width along this transverse slice is shown in the upper panel to the left.}
\end{figure}

The linearly polarized intensity present in our stacked radio map highlights the complex nature of the magnetized plasma within 3C 279 on parsec scales.  The observed peak in the polarized intensity (located to the left of the jet axis in Figure \ref{fig10}) may be associated with a standing feature within the jet downstream of the radio core.  This feature might represent a ``bend'' in the jet, which would create a localized ``hot spot'' resulting from a standing oblique shock that deflects the radio jet as it encounters the surrounding ambient medium.  The fact that the polarimetric signature of the sheath is confined to such a small region is encouraging from a modeling standpoint.  The \textit{Ring of Fire} model would predict that, as a blob propagates through this particular region within 3C 279, one might expect to observe an orphan flare due to the inverse-Compton scattering of this segment of sheath's photons off of the blob's electrons.  The VLBA-BU-Blazar program detected the ejection of a blob 57 $\pm$ 47 days before this orphan flare with a bulk Lorentz factor of $\Gamma \sim 21$ (Jorstad - private communication).  Perhaps the orphan flare in Figure \ref{fig11} is associated with this blob as it propagates through the sheath region highlighted downstream of the core in Figure \ref{fig10}.  A second blob was also detected 196 $\pm$ 33 days after this flare with a bulk Lorentz factor of  $\Gamma \sim 31$. 

\begin{figure*}
  \setlength{\abovecaptionskip}{-6pt}
  \begin{center}
    \scalebox{0.95}{\includegraphics[width=2.0\columnwidth,clip]{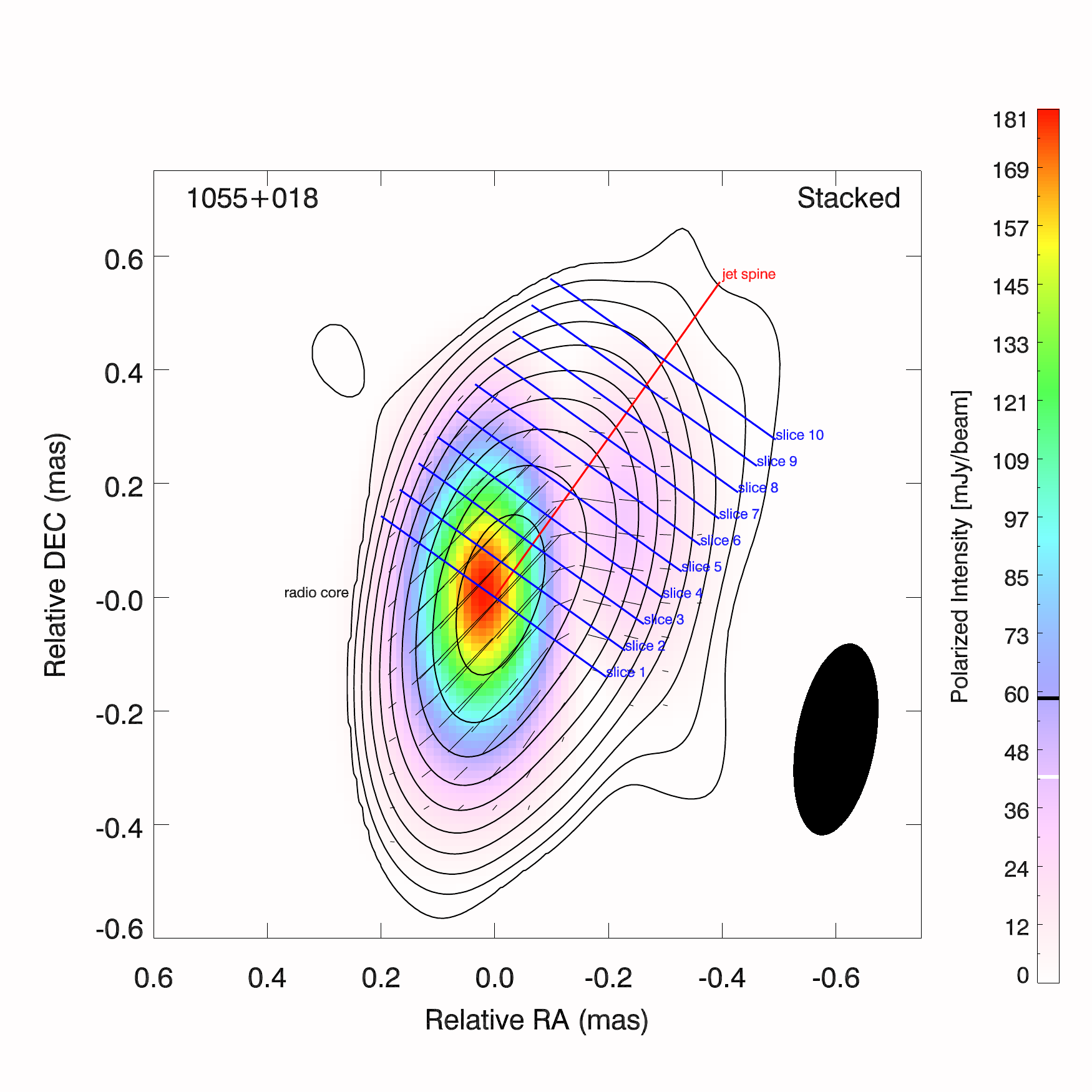}}
  \end{center}
  \caption{\label{fig13}A stacked map of 43 GHz images of 1055$+$018 spanning twenty epochs of observation from 2008 to 2016.  The black contours correspond to total intensity (contour levels are: 2.8, 5.7, 11.4, 22.9, 45.8, 91.6, 183.3, 366.7, 733.4, 1466.9 mJy beam$^{-1}$), whereas the underlying color scheme corresponds to polarized intensity (see color bar to the right for the flux levels), with the EVPAs denoted by black line segments.  The EVPAs indicate the orientation of linear polarization as projected onto the plane of the sky.   All images have been convolved with a Gaussian beam, shown in the bottom right corner of the stacked map.  A jet spine is plotted in red based on the transverse symmetry of the jet, across which ten transverse slices through the data are taken (shown in blue).  The profiles of the emission parameters along the third slice are shown in Figure \ref{fig15}.}
\end{figure*}

\newpage

\subsection{\rm{1055$+$018}}

Figure \ref{fig13} illustrates a stacked radio map of the blazar 1055$+$018.  In this stacked map the peak of the polarized intensity is co-spatial with the radio core.  The EVPAs within the radio core are roughly aligned with the axis of the jet (in red) indicative of shock acceleration compressing and aligning the magnetic field perpendicular to the jet flow.  Slightly downstream of the polarized emission in the core is a second polarimetric feature on the edge of the jet (to the west).  Within this second feature the EVPAs are inclined to the jet axis shown in red.  This feature is not mirrored on the other side of the jet as in the map of 4C 71$.$07 (Figure \ref{fig6}).  The sheath polarization that we do detect is probably Doppler beamed to some extent.  It is possible that there is a sheath located on the far side of the jet (to the east) that is undetected due to weaker Doppler beaming as a result of jet orientation.

\begin{figure}
\epsscale{1.15}
\plotone{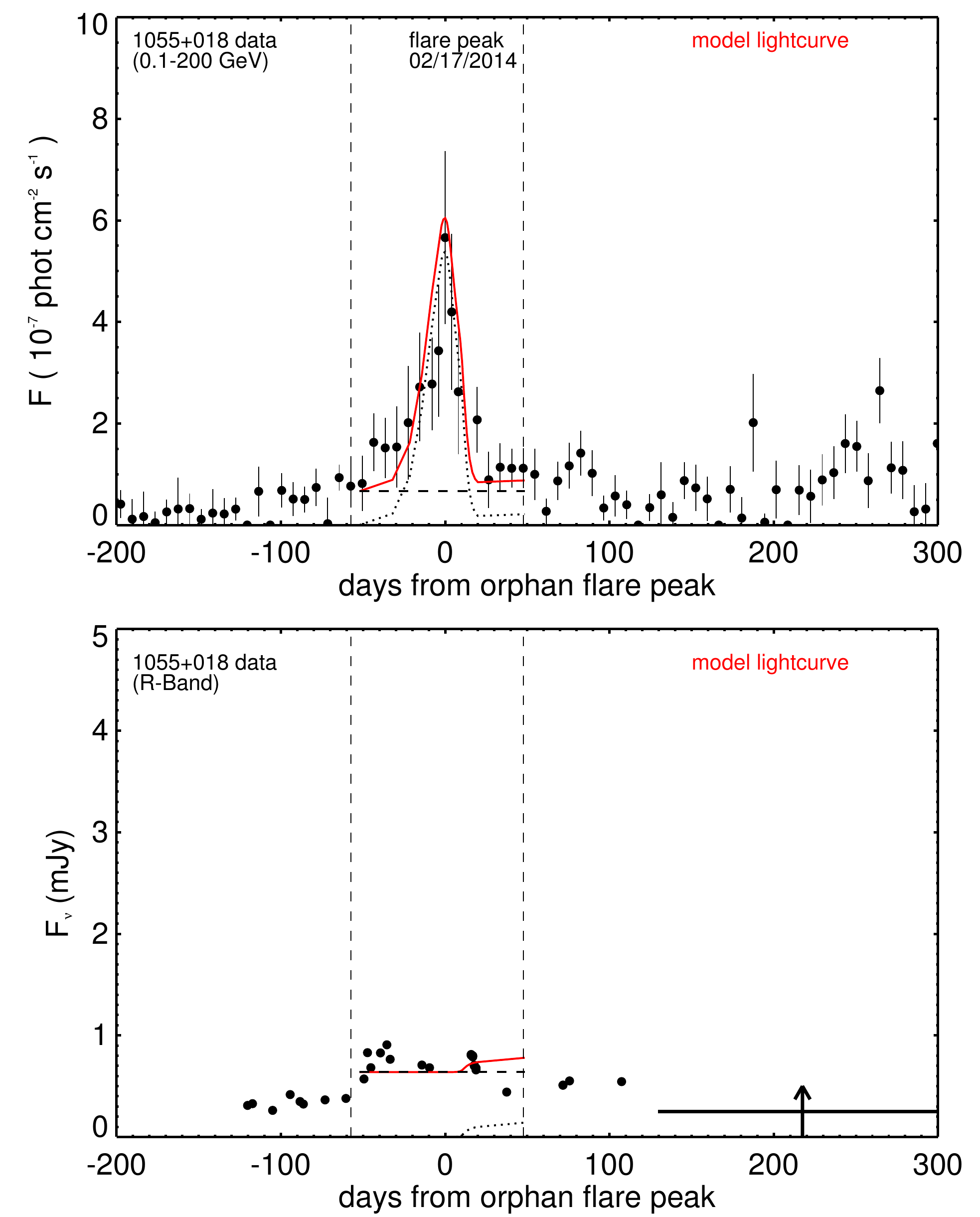}
\caption{\label{fig14}Light curves of 1055$+$018 (black circles) in the $\gamma$-rays (upper panel) and optical (lower panel).  The model light curves are overlaid in red and consist of the superposition of the model baseline level of flux in each band (the horizontal dashed black line) and the variable emission produced by the model (the dotted black line).  The vertical arrow in the lower panel marks the time when a superluminal knot passed through the 43 GHz core of 1055$+$018, with the horizontal bar representing the uncertainty in this time.}
\end{figure}

The Fermi LAT detected an orphan $\gamma$-ray flare with no optical counterpart in 1055$+$018 during February 2014.  The corresponding $\gamma$-ray and optical light curves are shown in Figure \ref{fig14}.  Table \ref{tab2} lists the \textit{Ring of Fire} model parameters used to obtain a fit to this orphan $\gamma$-ray flare, with the model light curves overlaid in red.  The nature of the blob's acceleration was adjusted until a fit to this asymmetric orphan flare profile (with a more gradual flare onset and a steeper flare decay) was obtained.  As with the other orphan flare models presented in this paper, the ring parameters were held constant and the blob's injection power was altered to fit the corresponding level of flux.  There is a slight ``bump'' in the synthetic optical light curve due to the Doppler boost of the emission gained from the blob's acceleration down the jet spine.  The synchrotron emission in the co-moving frame of the blob, however, remains constant due to the fixed nature of the magnetic field within the blob during the simulation.  
The VLBA-BU-Blazar program detected the ejection of a blob from the radio core of 1055$+$018 214 $\pm$ 88 days after this flare with a bulk Lorentz factor of $\Gamma \sim 15$ (Jorstad - private communication).  This VLBI speed is similar to our model value of $\Gamma_{\rm final} \sim 20$ \linebreak (see Table \ref{tab2}).  The \textit{Ring of Fire} posits that the site of orphan $\gamma$-ray flare production is located upstream of the radio core.  Our model, therefore, predicts that one should indeed see radio blobs pass through the 43 GHz core after the occurrence of an orphan flare.       

\begin{figure}
  \setlength{\abovecaptionskip}{-6pt}
  \begin{center}
  \hspace*{-2.2cm}
    \scalebox{0.93}{\includegraphics[width=1.4\columnwidth,clip]{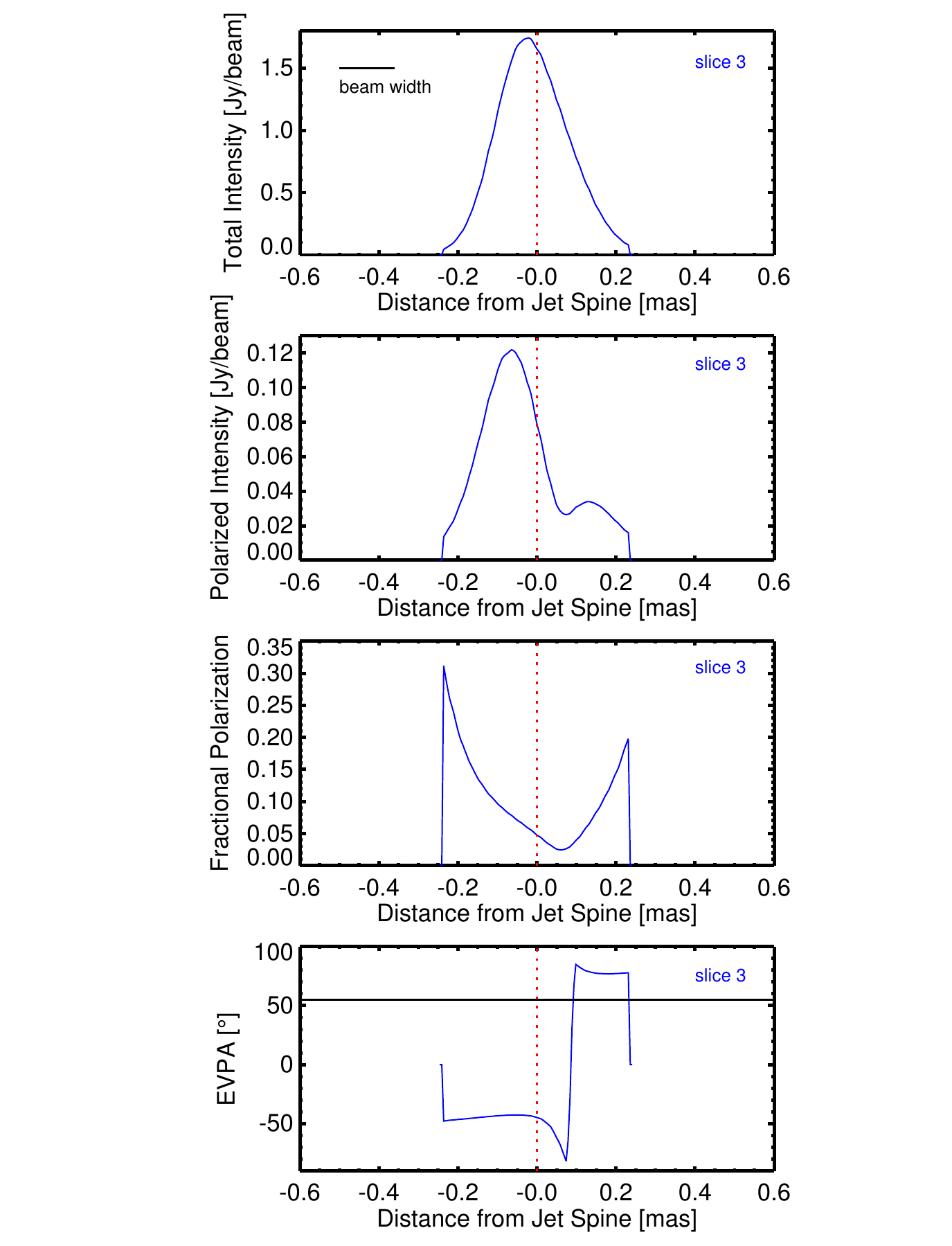}}
  \end{center}
  \caption{\label{fig15}Variations of the emission parameters of 1055$+$018 transverse to the jet axis (dashed red line) for slice 3 (shown in Figure \ref{fig13}) from top to bottom: total intensity, polarized intensity, fractional polarization, and EVPA.  The fractional polarization increasing toward the edges of the jet is a predicted polarimetric signature of a jet sheath (as discussed in \S1), with the EVPA transverse to the axis.  The black line in the lower panel delineates EVPAs that are exactly $\perp$ to the jet spine. The beam width along this transverse slice is shown in the upper panel to the left.}
\end{figure}

The profiles of the Stokes emission parameters transverse to the jet axis along the third slice through our stacked map of 1055$+$018 are shown in Figure \ref{fig15}.  The increase in fractional linear polarization towards the edges of the jet is clearly visible in the lower middle panel.  The linearly polarized intensity profile (upper middle panel) is asymmetric across the jet axis.  

\subsubsection{}
\label{stuff}

It should be pointed out that an alternate interpretation of the polarimetric profiles presented in Figure \ref{fig15} (and indeed the rest of the profiles presented in this paper) is that, in contrast to the spine-sheath model, the jet carries a large scale helical magnetic field (see, e.g., \citealt{gabuzda13}).  A determination of the rotation measure (RM) across the width of the jet can, in theory, distinguish between these two scenarios.  We adopt the spine-sheath interpretation in this paper.  Regardless of whether the polarization that we detect emanates from jet shear or a large scale helical magnetic field, this polarization does delineate a distinct region of magnetized plasma that is located in the more slowly moving periphery of the jet relative to the spine. This plasma will produce a localized source of seed photons (required by the \textit{Ring of Fire} model).     

\begin{figure*}
  \setlength{\abovecaptionskip}{-6pt}
  \begin{center}
    \scalebox{0.99}{\includegraphics[width=2.0\columnwidth,clip]{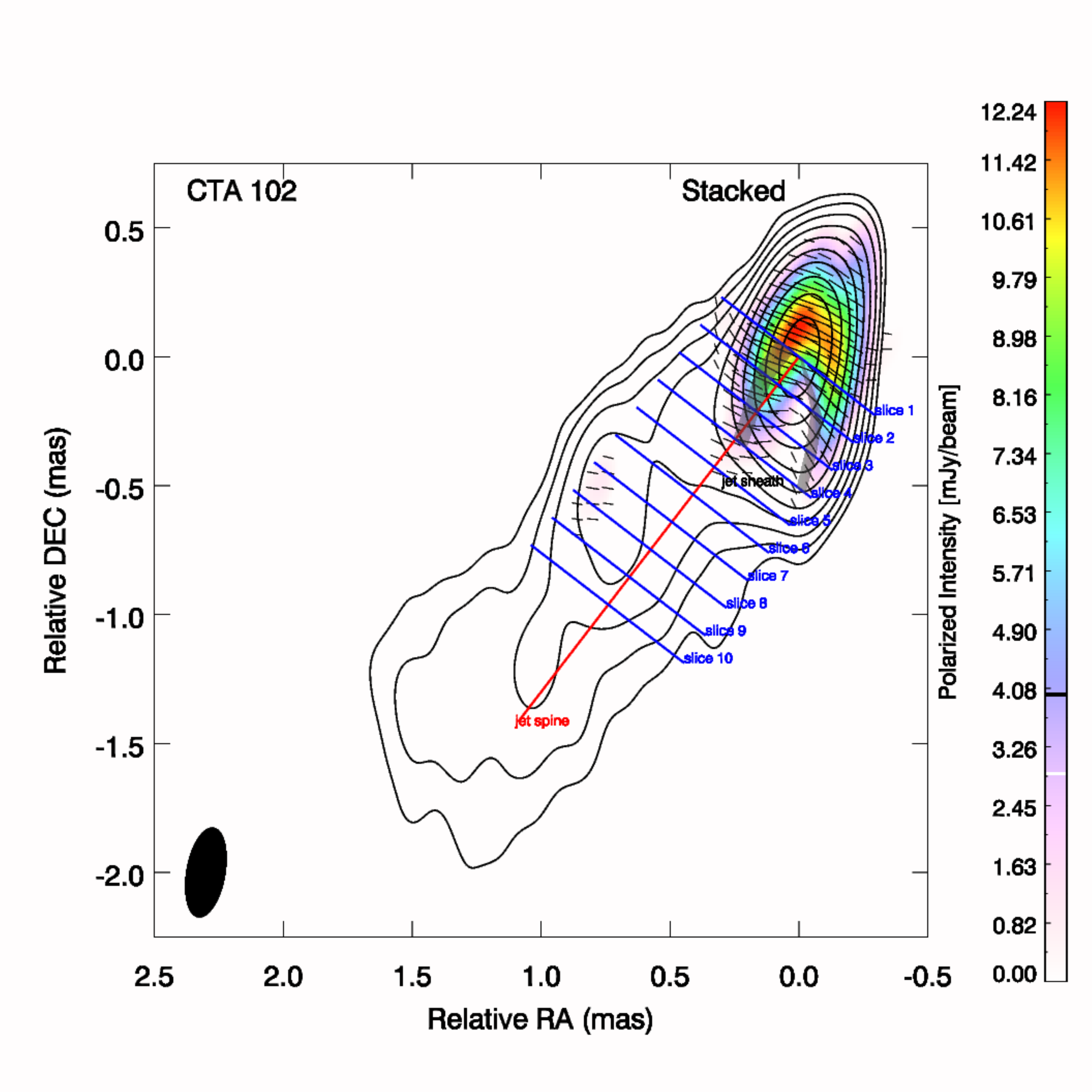}}
  \end{center}
  \vspace*{-0.1cm}
  \caption{\label{fig16}A stacked map of 43 GHz images of CTA 102 spanning twenty epochs of observation from 2008 to 2016.  The black contours correspond to total intensity (contour levels are: 2.0, 4.0, 8.0, 16.1, 32.2, 64.5, 129.0, 258.1, 516.3, 1032.6 mJy beam$^{-1}$), whereas the underlying color scheme corresponds to polarized intensity (see color bar to the right for the flux levels), with the EVPAs denoted by black line segments.  The EVPAs indicate the orientation of linear polarization as projected onto the plane of the sky.   All images have been convolved with a Gaussian beam, shown in the bottom left corner of the stacked map.  A jet spine is plotted in red based on the transverse symmetry of the jet, across which ten transverse slices through the data are taken (shown in blue).  The profiles of the emission parameters along the first slice are shown in Figure \ref{fig18}.  The shaded gray regions in the above map highlight the nominal location of the jet sheath of CTA 102 as determined by the procedure illustrated in Figure \ref{fig19}.}
\end{figure*}

\newpage

\subsection{\rm{CTA 102}}

In Figure \ref{fig16} we present a stacked radio map of the blazar CTA 102.  Similar to 4C 71$.$07, a continuous polarimetric signature of a jet sheath is evident down the edges of the jet in this stacked radio image.  Given the continuous nature of the jet sheath present, we again compute an estimate of the bolometric luminosity of the sheath.  The extent of the polarized intensity profile along each of the slices was used to trace the shaded gray regions shown in Figure \ref{fig16}.  This was done by using the width of the sheath on each side of each slice (shown explicitly in Figure \ref{fig19}) after determining, arbitrarily, the location where the polarized flux falls below 0.85 times the peak value.  An estimate of the bolometric luminosity ($L_{\rm bol}$) of the sheath is computed by adding all of the deconvolved `sheath' flux (see Figures \ref{fig18} \& \ref{fig19}) contained within the gray shaded regions of the stacked map above.  As in \S4.2, we convert this inferred sheath flux at 43 GHz into a spectral luminosity and assume a power-law ($L_{\nu} \propto \nu^{-\alpha}$), which we then integrate from $\nu_{\rm min} = 10^{9} ~ \rm Hz$ to $\nu_{\rm max} = 5 \times 10^{13} ~ \rm Hz$.  \linebreak Using Equation 1, we obtain: $ L_{\rm bol} \sim 1 \times 10^{45} ~ \rm erg ~ \rm s^{-1}$, where we have again assumed a spectral index of $\alpha \sim 1.0$.  Similar to 4C 71$.$07, our estimate above indicates that the jet sheath in CTA 102 is an important source of seed photons even at parsec scales. We again, however, point out that convolution with the beam adds a contribution of flux from the spine to this sheath flux estimate due to our limited resolution transverse to the jet axis. An estimate of the inner ($r_{\rm min}$) and outer ($r_{\rm max}$) radii of the sheath was also obtained:  we find that $r_{\rm min} \sim 0.84 ~ \rm pc$ and $r_{\rm max} \sim 1.28 ~ \rm pc$.  Similar to the sheath in 4C 71$.$07, these values are much larger than the inner and outer radii used in our model of the ring (see Table \ref{tab1}) that we posit to exist upstream of the radio core.

\begin{figure}
\epsscale{1.0}
\plotone{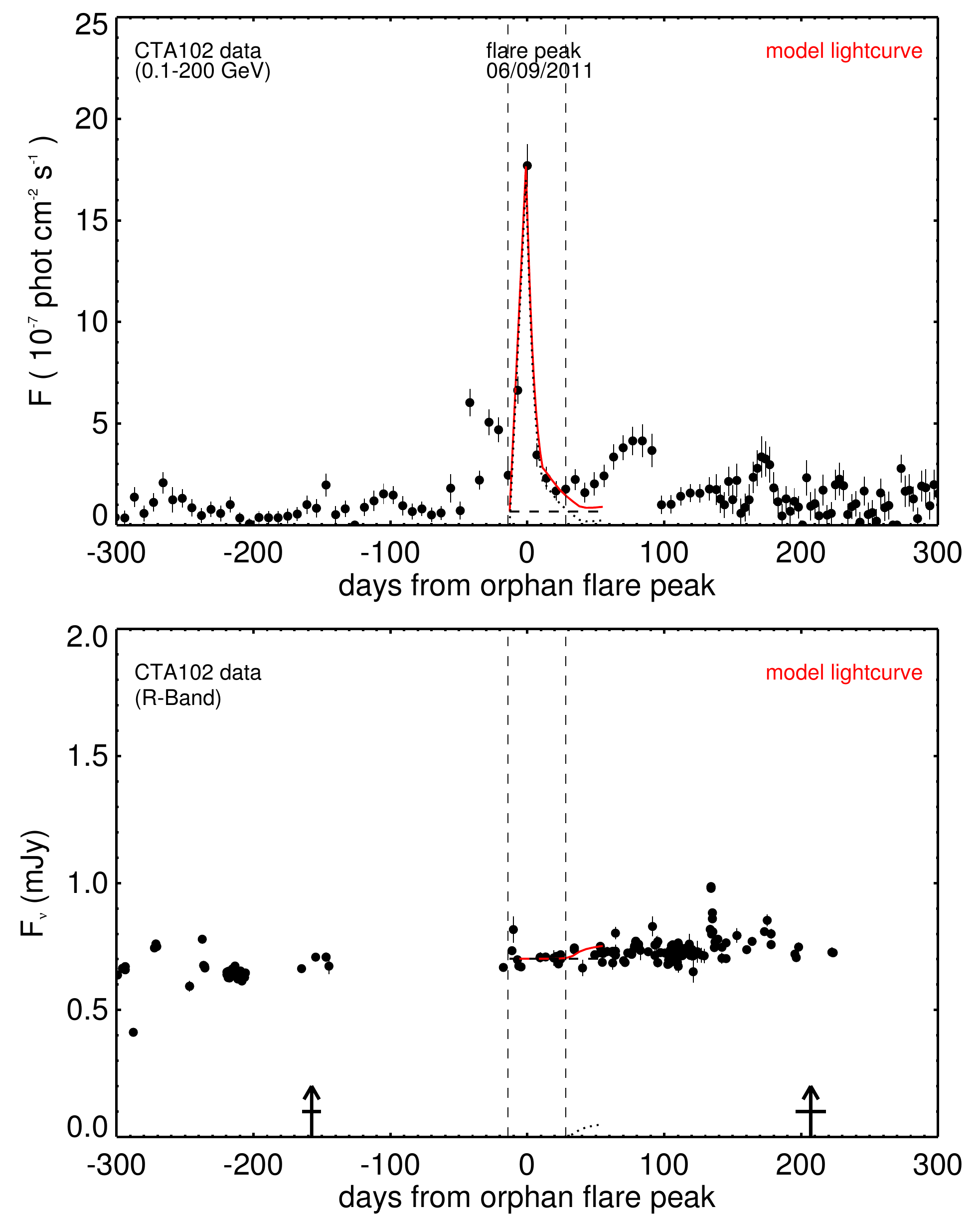}
\caption{\label{fig17}Light curves of CTA 102 (black circles) in the $\gamma$-rays (upper panel) and optical (lower panel).  The model light curves are overlaid in red and consist of the superposition of the model baseline level of flux in each band (the horizontal dashed black line) and the variable emission produced by the model (the dotted black line).  The vertical arrows in the lower panel mark the time when superluminal knots passed through the 43 GHz core of CTA 102, with the horizontal bars representing the uncertainty in these times.}
\end{figure}  

The Fermi LAT detected an orphan $\gamma$-ray flare from CTA 102 during June 2011 (Figure \ref{fig17}).  Table \ref{tab2} lists the \textit{Ring of Fire} model parameters used to obtain a fit to this orphan flare.  The VLBA also observed the ejection of two blobs from the radio core of CTA 102, one occurring roughly 200 days before the onset of the orphan flare and the other roughly 200 days after. The times of the blob ejections are demarcated by vertical arrows in the lower panel of Figure \ref{fig17}.  \cite{casadio15} performed a detailed analysis of the kinematics of the blobs detected by the VLBA during these epochs and reported bulk Lorentz factors of the order $\Gamma \sim 26$.  These VLBI speeds are not all that dissimilar from the blob speed used in our model of this orphan flare ($\Gamma_{\rm final} \sim 35$).  The blob that passed through the core of CTA 102 roughly 200 days after this orphan flare may have supplied the electrons required to scatter the sheath photons.

\begin{figure}
  \setlength{\abovecaptionskip}{-6pt}
  \begin{center}
  \hspace*{-1.87cm}
    \scalebox{0.93}{\includegraphics[width=1.26\columnwidth,clip]{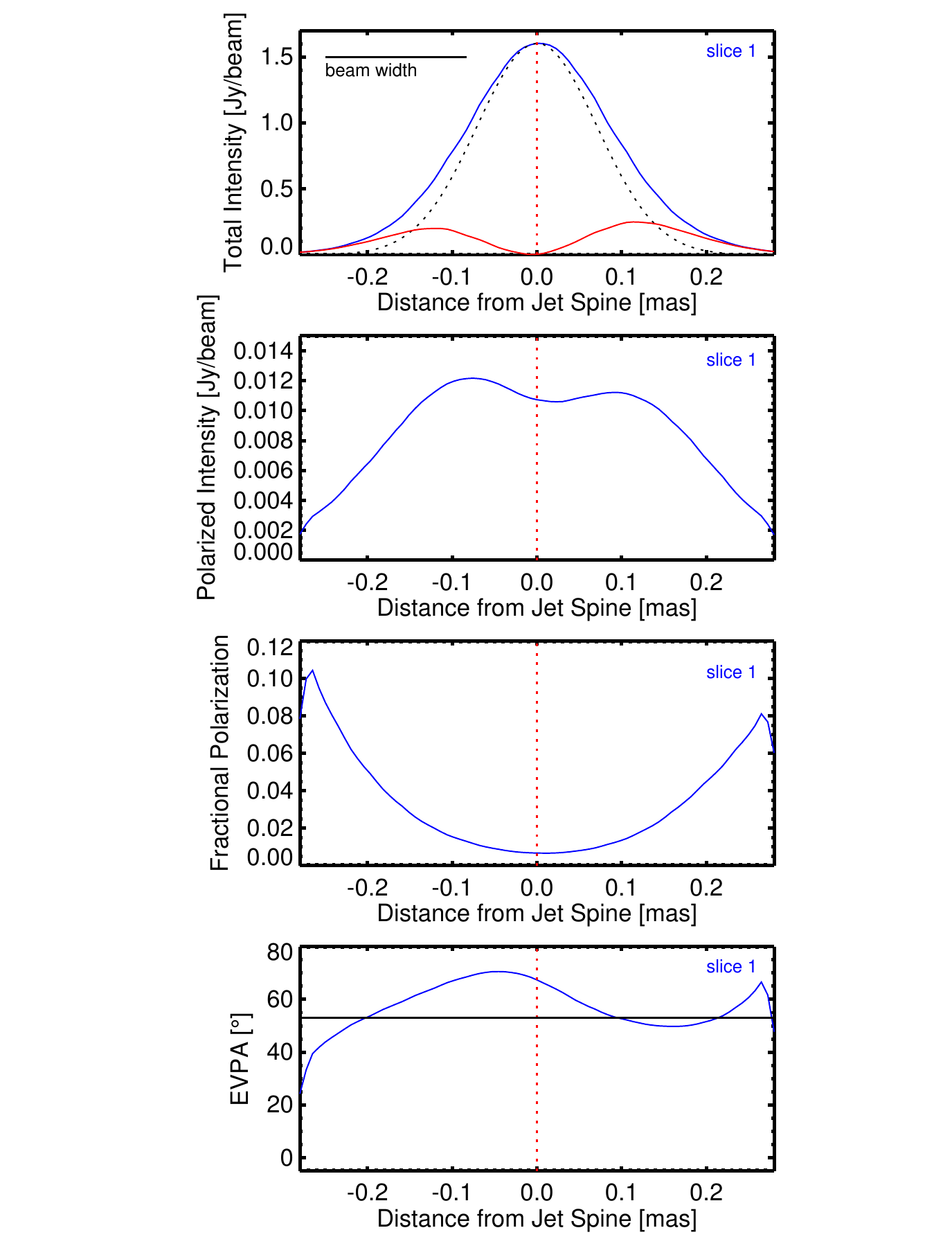}}
  \end{center}
  \vspace*{-0.2cm}
  \caption{\label{fig18}Variations of the emission parameters (solid blue lines) transverse to the jet axis (dashed red line) of CTA 102 for slice 1 (shown in Figure \ref{fig16}) from top to bottom: total intensity, polarized intensity, fractional polarization, and EVPA.  The fractional polarization increasing toward the edges of the jet is a predicted polarimetric signature of a jet sheath (as discussed in \S1), with the EVPA transverse to the axis.  The black line in the lower panel delineates EVPAs that are exactly $\perp$ to the jet spine. The beam width along this transverse slice is shown in the upper panel to the left. The beam profile (dashed black line) is deconvolved from the total intensity profile to estimate a `sheath' total intensity profile (solid red line).}
\end{figure}

\begin{figure}
  \setlength{\abovecaptionskip}{-6pt}
  \begin{center}
  \hspace*{0.36cm}
    \scalebox{0.97}{\includegraphics[width=0.82\columnwidth,clip]{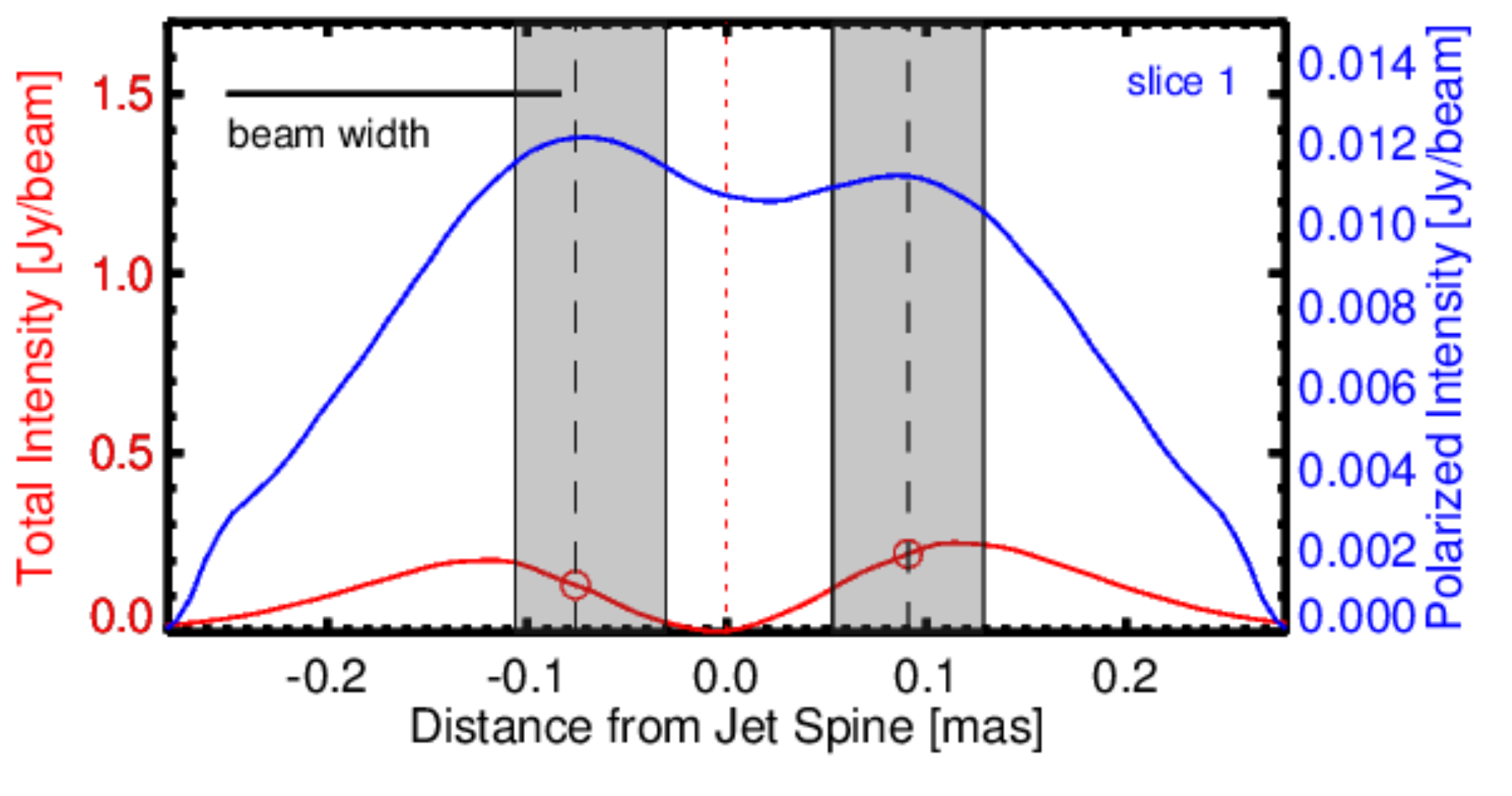}}
  \end{center}
  \vspace*{-0.3cm}
  \caption{\label{fig19}The polarized intensity profile of CTA 102 along slice 1 (solid blue line corresponding to the right-hand axis) is overlaid upon the deconvolved `sheath' total intensity profile for slice 1 (solid red line corresponding to the left-hand axis).  We use this double-peaked profile to highlight the likely location of the sheath within each slice (shaded gray regions). The sheath's contribution to the jet's total intensity profile is estimated by the values of total intensity (red circles) that are cospatial with the peaks of the polarized intensity profile (demarcated by dashed vertical lines). This procedure is then repeated for each slice, thus tracing out the sheath (shaded gray regions) shown in Figure \ref{fig16}.}
\end{figure} 

\begin{figure*}
  \setlength{\abovecaptionskip}{-6pt}
  \begin{center}
    \scalebox{0.96}{\includegraphics[width=2.0\columnwidth,clip]{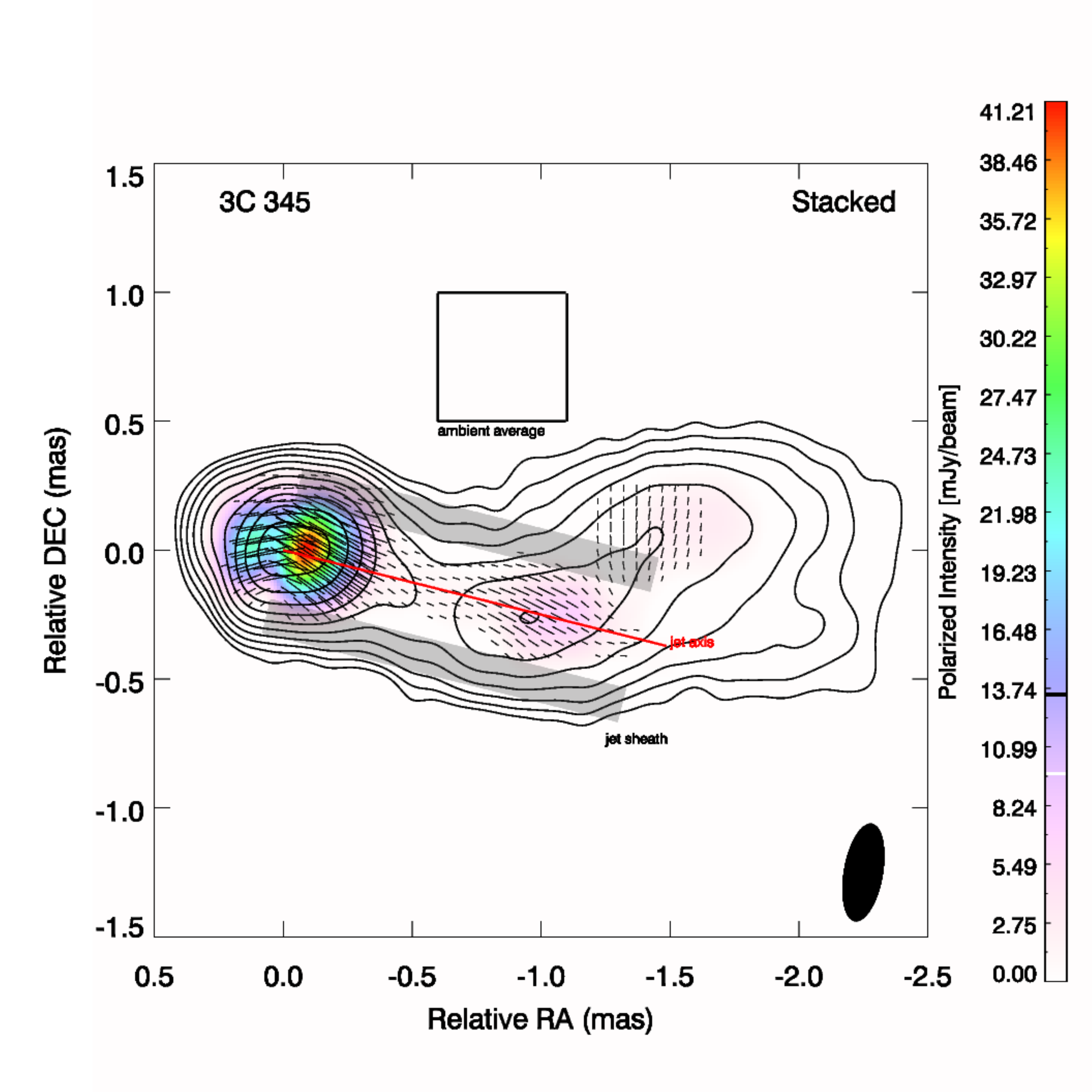}}
  \end{center}
  \vspace*{-0.1cm}
  \caption{\label{fig20}A stacked map of 43 GHz images of 3C 345 spanning twenty epochs of observation from 2008 to 2016.  The black contours correspond to total intensity (contour levels are: 1.2, 2.4, 4.9, 9.9, 19.9, 39.9, 79.9, 159.8, 319.6, 639.2 mJy beam$^{-1}$), whereas the underlying color scheme corresponds to polarized intensity (see color bar to the right for the flux levels), with the EVPAs denoted by black line segments.  The EVPAs indicate the orientation of linear polarization as projected onto the plane of the sky.   All images have been convolved with a Gaussian beam, shown in the bottom right corner of the stacked map.  The shaded gray regions in the above map highlight the absence of the polarimetric signature of a jet sheath within 3C 345.  The box delineates a region within which we compute an estimate of the ambient noise in this stacked radio map.}
\end{figure*}

\newpage

\subsection{\rm{3C 345}}

\begin{figure}
\epsscale{1.0}
\plotone{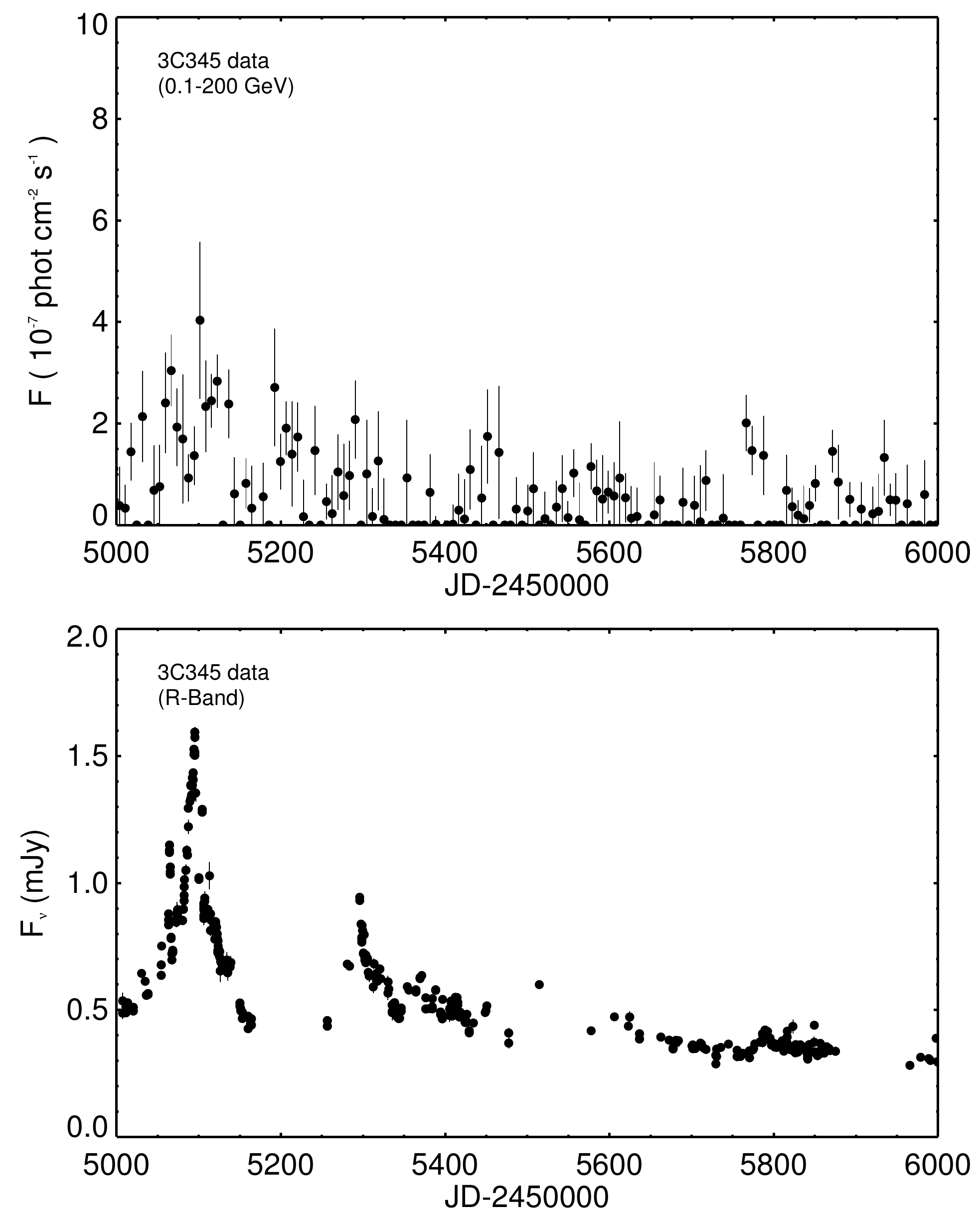}
\caption{\label{fig21}Light curves of 3C 345 (black circles) in the $\gamma$-rays (upper panel) and optical (lower panel).  In contrast to the orphan flare profiles presented in sections \S4.1-\S4.5, here there is little evidence of prominent orphan $\gamma$-ray flaring activity.}
\end{figure}

We display a stacked radio map of the blazar 3C 345 in Figure \ref{fig20}.  In contrast to the blazars discussed above, there is no polarimetric signature of a jet sheath present in our stacked map.  Instead, the polarization is concentrated along the center of the jet, with the EVPAs aligned predominantly parallel to the jet axis.  This alignment, in contrast to jet shear, is that expected from diffusive shock acceleration within the jet that compresses the magnetic field and aligns the field lines perpendicular to the jet flow, resulting in EVPAs oriented parallel to the spine of the jet (\citealt{wardle94}).  The jet undergoes two bends downstream of the radio core, and the EVPAs maintain their orientation with respect to the jet axis around each bend.  We searched through archival $\gamma$-ray and optical light curves to look for instances of orphan $\gamma$-ray flares within 3C 345.  We were unable to find any (see Figure \ref{fig21}).  This suggests that an enhancement in the jet sheath is not present within 3C 345.  The lack of a polarimetric signature of a jet sheath within 3C 345 might indicate that, while jet sheaths are almost certainly present within all blazars, the sheath itself only becomes ``enhanced'' at certain times within the span of a blazar's lifetime (perhaps as a portion of the sheath encounters a standing shock within the jet).  This would explain why the majority of blazar $\gamma$-ray flares are not orphan in nature.  It is also possible that the jet sheath is present, but we are unable to detect it due to the extreme gradient in intrinsic brightness between the spine and the sheath.  In order to compute a rough estimate of the upper limit of the bolometric luminosity of a jet sheath within 3C 345, we trace (in gray) a region on the periphery of the jet within which a sheath might exist (see Figure \ref{fig20}).  Similar to \S4.2, we add up the flux in these gray sheath regions at 43 GHz, but instead of using jet flux we use an estimate of the ambient noise in our stacked map (obtained within the box shown in Figure \ref{fig20}).  Using Equation 1, we obtain an upper limit to the bolometric luminosity of a sheath of $L_{\rm bol} \sim 5 \times 10^{42} ~ \rm erg ~ \rm s^{-1}$, where we have again assumed a spectral index of $\alpha \sim 1.0$ and limits of integration of $\nu_{\rm min} = 10^{9} ~ \rm Hz$ and $\nu_{\rm max} = 5 \times 10^{13} ~ \rm Hz$.  The majority of the stacked maps presented in this paper are comprised of twenty epochs of observation.  It might be profitable to continue stacking to see whether additional maps yield a still fainter signature of a jet sheath within 3C 345.

\section{Summary and Conclusions}

We have created stacked 43 GHz maps of the  milliarcsecond-scale structure of the total and polarized intensity of the blazars 3C 273, 4C 71$.$07, 3C 279, 1055$+$018, CTA 102, and 3C 345.  Each stacked map is carefully constructed from data collected over the course of eight years as part of the VLBA-BU-BLAZAR program, creating images with unprecedented dynamic range, sensitivity, and angular resolution.  We find the polarimetric signature of a jet sheath within five of these stacked radio maps.  In the stacked radio map within which we did not detect the polarimetric signature of a jet sheath (3C 345 - Figure \ref{fig20}), we were also unable to find any instances of orphan $\gamma$-ray flares.  This perhaps indicates that localized enhancements within the jet sheath, required by the \textit{Ring of Fire} model, persist only for a short period of time.  This would explain why the majority of $\gamma$-ray flares detected within blazars are not orphan in nature.  The five blazars (3C 273, 4C 71$.$07, 3C 279, 1055$+$018, and CTA 102) in which we detect a jet sheath have all exhibited orphan flaring behavior.  

The \textit{Ring of Fire} model, developed by \cite{macdonald15}, is able to reproduce the basic time profiles of these orphan flares.  This model invokes the presence of a localized enhancement within the jet sheath (a shocked segment/ring) to create a source of seed photons that are inverse-Compton scattered up to high energies by electrons contained within a blob moving relativistically along the jet spine and through the enhancement.  By simply varying the nature of the blob's acceleration down the spine of the jet, we are able to obtain fits to all five of these orphan flare profiles.  This highlights the potential of the \textit{Ring of Fire} model to explain all orphan $\gamma$-ray flaring behavior from within blazars based on the radiative interplay between plasma sheaths and plasma blobs within these relativistic jets.  As discussed in \cite{macdonald15}, we find no conflict (as suggested by \citealt{nalewajko14}) in the sheath's ability to supply sufficient levels of seed photons for inverse-Compton scattering while not simultaneously outshining the jet spine.  The six orphan flares we present in this paper were contemporaneous with radio blob ejections detected by analysis of VLBA images (Figures \ref{fig4}, \ref{fig7}, \ref{fig11}, \ref{fig14}, \& \ref{fig17}), thus lending further observational support to the plausibility of our model.  A rough estimate of the location(s) in the jet of the enhancement in the photon field of the sheath (i.e., the ring) can be obtained from the time ($\Delta t$) between the respective orphan flare peaks and when each blob passed through the radio core.  If we assume, for simplicity, that the blob travels with a constant velocity ($\beta_{\rm{app}}$) along the jet, then the distance $\Delta z$ upstream/downstream of the flare from the radio core is given by
\begin{equation}\label{eq_stream}
\Delta z \simeq \frac{ \beta_{ \rm{app} } ~ \Delta t ~ c }{ \rm{sin}(\theta_{\rm obs}) } ~ .
\end{equation}
The values obtained from Equation \ref{eq_stream} for the orphan flares presented in \S\S4.1-4.5 are listed in Table \ref{tab2}.  Our model allows for a different location of $\gamma$-ray flare production within the jet than those considered in most interpretations of blazar light curves, namely, the BLR, a dusty torus, or a standing shock in the millimeter-wave ``core'' \citep[e.g.,][]{marscher14,joshi14}.

\subsection*{Acknowledgements}

Funding for this research was provided by a Canadian NSERC PGS D2 Doctoral Fellowship and by NASA Fermi Guest Investigator grants NNX11AQ03G, NNX12AO79G, NNX12AO59G and NNX14AQ58G.  The authors are grateful to Karen Williamson for compiling the multi-wavelength data used in Figures \ref{fig4}, \ref{fig7}, \ref{fig11}, \ref{fig14}, \ref{fig17}, and \ref{fig21} and to the anonymous referee for a thorough and helpful review. The VLBA is an instrument of the Long Baseline Observatory. The Long Baseline Observatory is a facility of the National Science Foundation operated by Associated Universities, Inc.

\begin{deluxetable*}{lccccc}
\tablecolumns{6}
\centering
\tablewidth{2.0\columnwidth}
\tabletypesize{\normalsize}
\tablecaption{Ring of Fire Source Parameters \label{tab2}}
\startdata
\tableline
\noalign{\smallskip}
\tableline
\noalign{\smallskip}
 & 3C 273 & 4C 71$.$07 & 3C 279 & 1055$+$018 & CTA 102 \\
\tableline
\noalign{\smallskip}
$\Gamma_{\rm{initial}}$ & 3.0 & 3.0 & 2.0 & 4.0 & 2.0 \\
\noalign{\smallskip}
$\Gamma_{\rm{final}}$ & 25.0 & 10.0 & 20.0 & 20.0 & 35.0 \\
\noalign{\smallskip}
$z_{\rm{initial}}$ (pc) & $-0.4$ & $-0.4$ & $-0.1$ & $-0.6$ & $-0.2$ \\
\noalign{\smallskip}
$z_{\rm{final}}$ (pc) & 0.0 & 0.0 & 0.1 & 0.16 & 0.0 \\
\noalign{\smallskip}
$P_{\rm{inj} ~ \rm{flare 1}}$ ($\rm{ergs} ~ s^{-1}$) & $5.0 \times 10^{45}$ & $1.0 \times 10^{45}$ & $3.0 \times 10^{44}$ & $3.0 \times 10^{44}$ & $4.6 \times 10^{45}$ \\
\noalign{\smallskip}
$P_{\rm{inj} ~ \rm{flare 2}}$ ($\rm{ergs} ~ s^{-1}$) & - & $9.0 \times 10^{44}$ & - & - & - \\
\noalign{\smallskip}
$\rm{Baseline} ~ \rm{Flux}_{~ \rm{optical}}$ (mJy) & 28.7 & 0.7 & 3.8 & 0.64 & 0.7 \\
\noalign{\smallskip}
$\rm{Baseline} ~ \rm{ Flux}_{~ \gamma-\rm{ray}}$ ($10^{-7} \rm{phot} ~ \rm{cm}^{-2} ~ \rm{s}^{-1}$) & 1.7 & 0.76 & 6.5 & 0.67 & 0.67 \\
\noalign{\smallskip}
$Z$ & 0.158 & 2.17 & 0.538 & 0.89 & 1.037 \\
\noalign{\smallskip}
$\theta_{\rm{obs}}$ & $6.1^{\circ}$ & $3.0^{\circ}$ & $2.1^{\circ}$ & $2.0^{\circ}$ & $2.6^{\circ}$ \\
\noalign{\smallskip}
$\Delta z$ (pc)            &    0.4               &   29.8             &   27.4              & 77.0               &   96.2 \\
\noalign{\smallskip}
relative to radio core & downstream  & upstream      &  downstream & upstream     & upstream
\enddata
\end{deluxetable*}

\newpage

\appendix
\section{\\A - Stacked Radio Map Time Sampling}

\begin{figure}[!htbp]
\begin{center}
\includegraphics[width=0.75\textwidth]{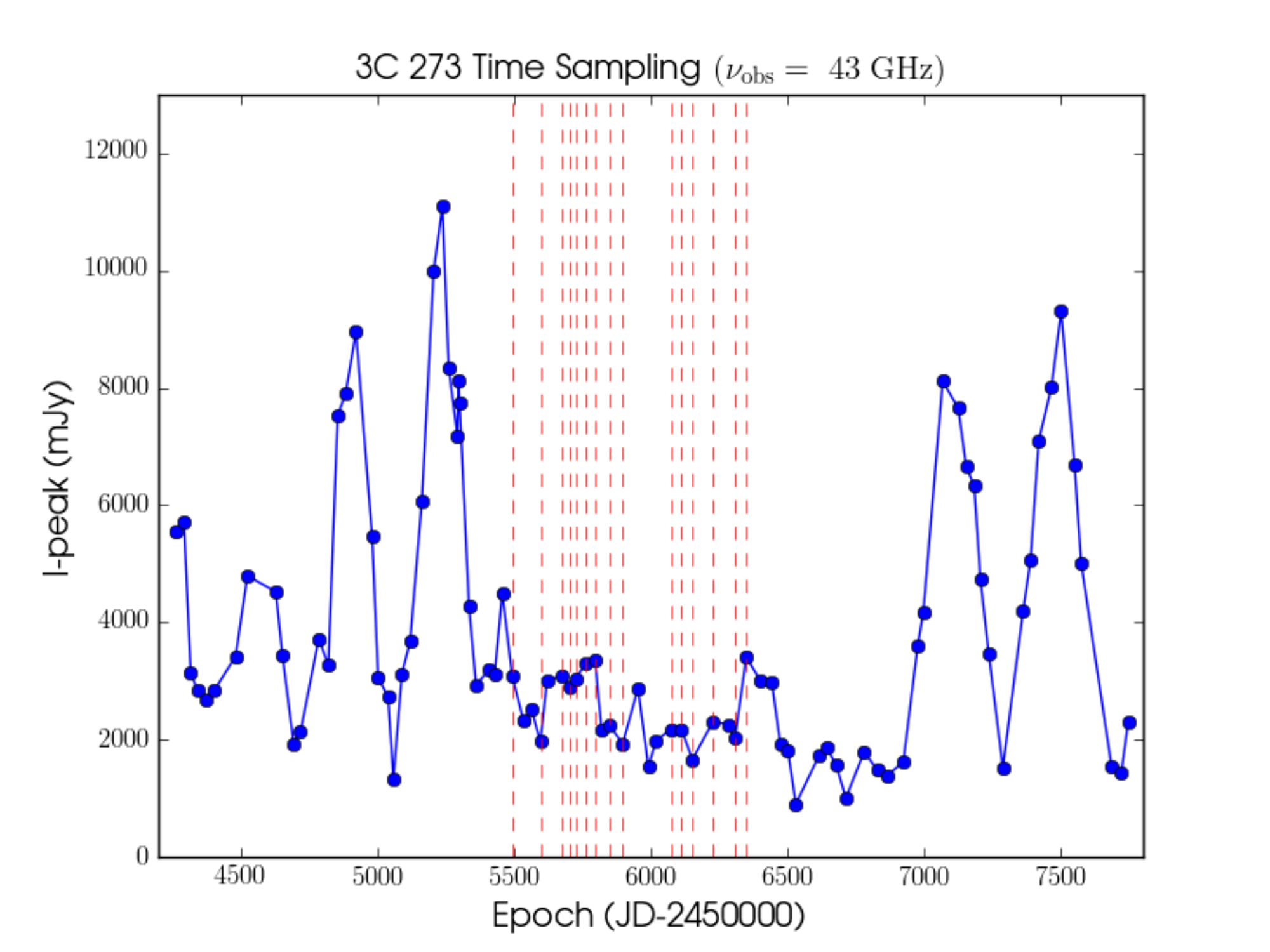}
\end{center}
\vspace{-0.3cm}
\caption{\label{fig22} Peak intensity at 43 GHz of 3C 273 versus time for the entire span of the VLBA-BU-BLAZAR program. The epochs included within the stacked image (Figure \ref{fig3}) are highlighted with dashed red vertical lines and correspond to periods of relatively low jet activity.}
\end{figure}

\begin{figure}[!htbp]
\begin{center}
\includegraphics[width=0.75\textwidth]{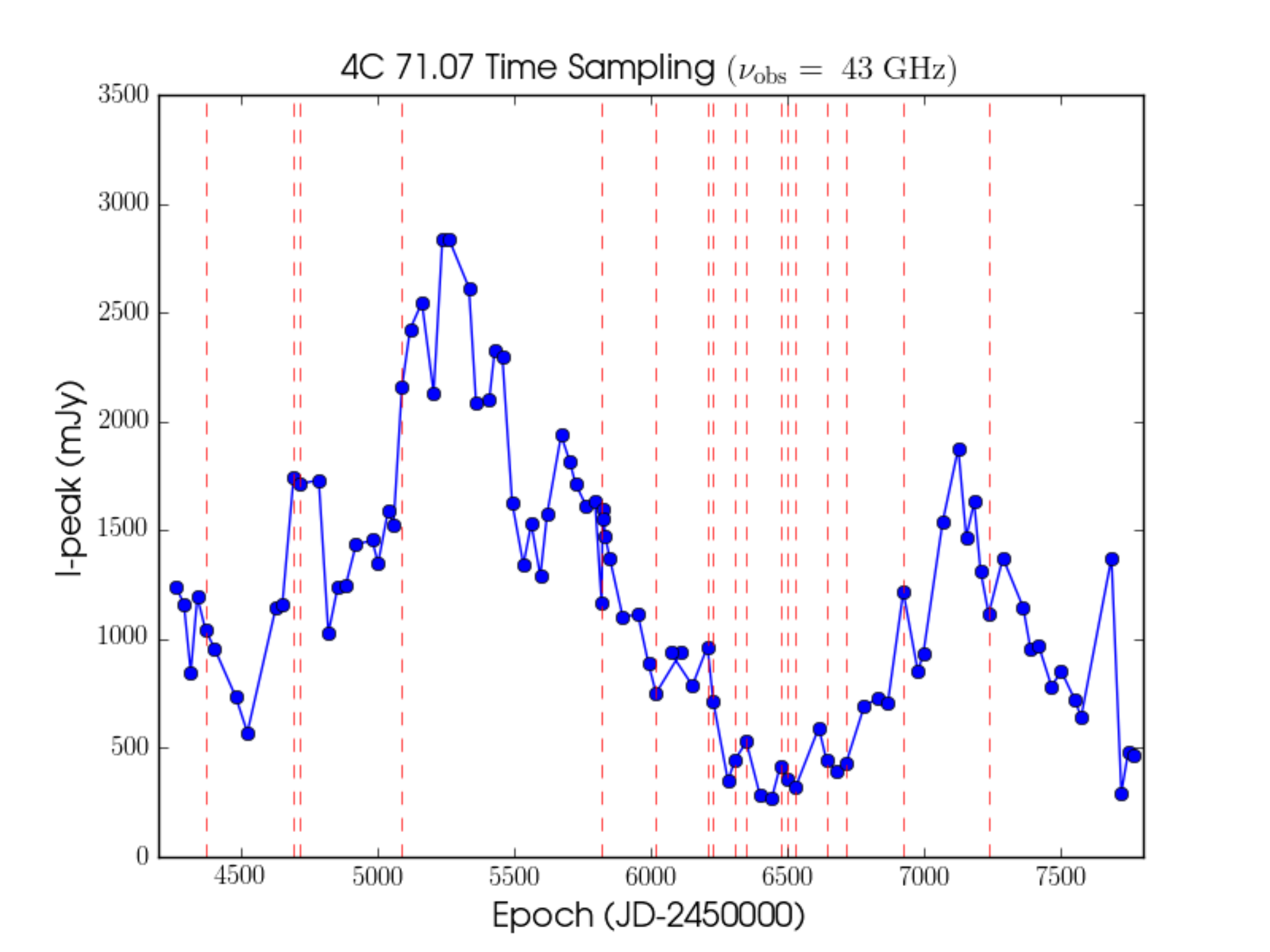}
\end{center}
\vspace{-0.3cm}
\caption{\label{fig23} Peak intensity at 43 GHz of 4C 71.07 versus time for the entire span of the VLBA-BU-BLAZAR program. The epochs included within the stacked image (Figure \ref{fig6}) are highlighted with dashed red vertical lines and correspond to periods of relatively low jet activity.}
\end{figure}

\begin{figure}[!htbp]
\begin{center}
\includegraphics[width=0.75\textwidth]{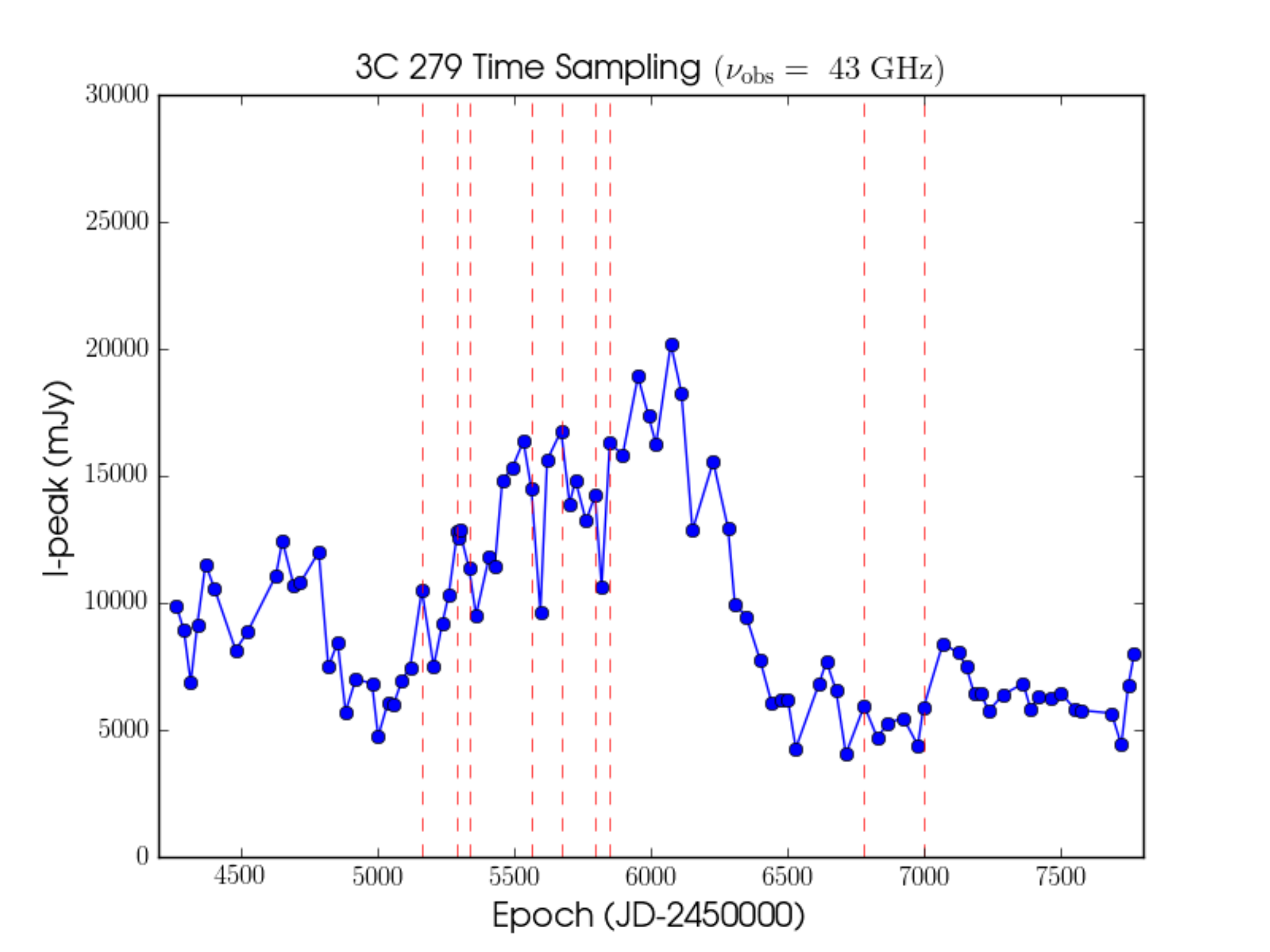}
\end{center}
\vspace{-0.3cm}
\caption{\label{fig24} Peak intensity at 43 GHz of 3C 279 versus time for the entire span of the VLBA-BU-BLAZAR program. The epochs included within the stacked image (Figure \ref{fig10}) are highlighted with dashed red vertical lines and correspond to periods of relatively low jet activity.}
\end{figure}

\begin{figure}[!htbp]
\begin{center}
\includegraphics[width=0.75\textwidth]{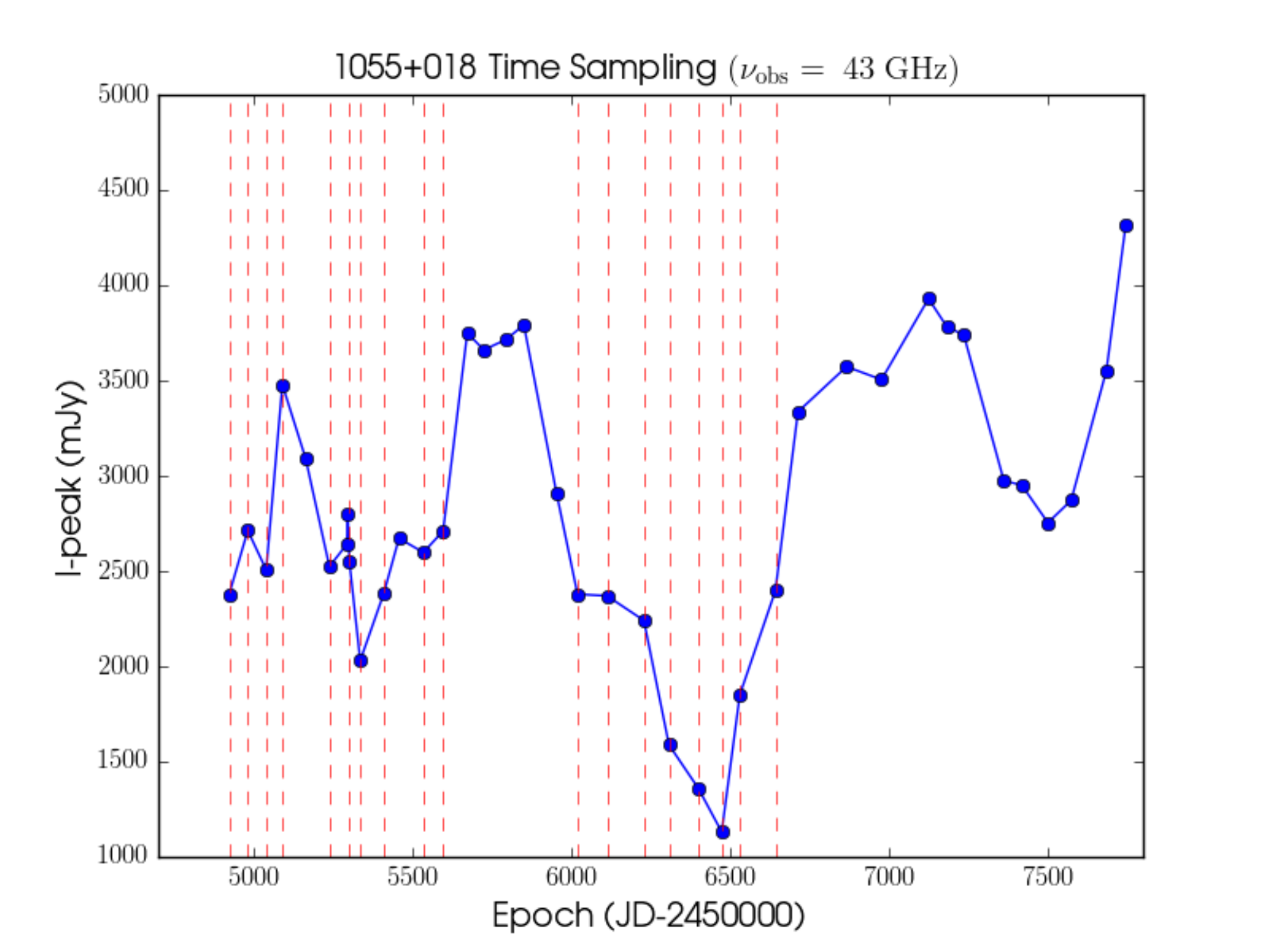}
\end{center}
\vspace{-0.3cm}
\caption{\label{fig25} Peak intensity at 43 GHz of 1055+018 versus time for the entire span of the VLBA-BU-BLAZAR program. The epochs included within the stacked image (Figure \ref{fig13}) are highlighted with dashed red vertical lines and correspond to periods of relatively low jet activity.}
\end{figure}

\begin{figure}[!htbp]
\begin{center}
\includegraphics[width=0.75\textwidth]{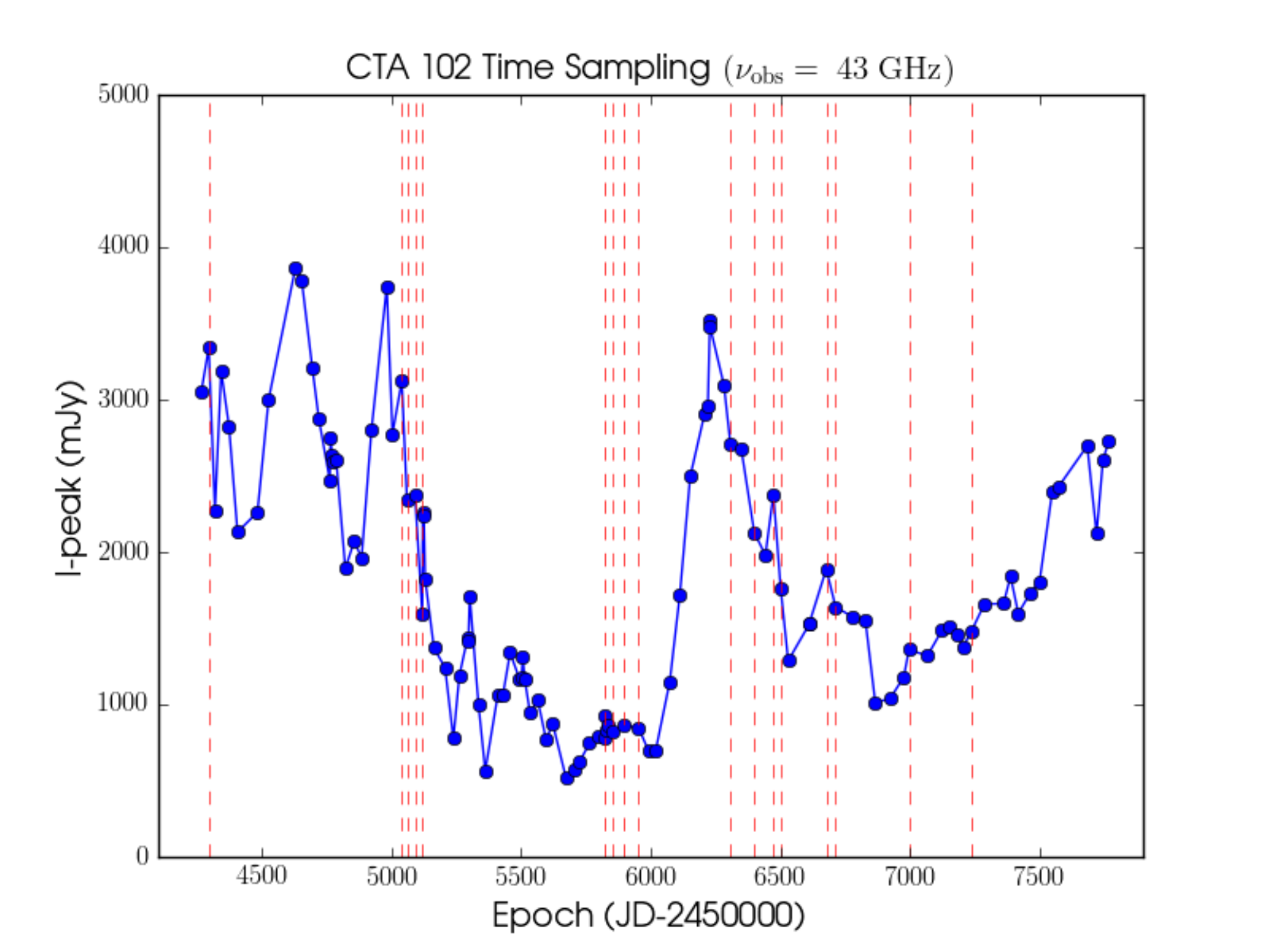}
\end{center}
\vspace{-0.3cm}
\caption{\label{fig26} Peak intensity at 43 GHz of CTA 102 versus time for the entire span of the VLBA-BU-BLAZAR program. The epochs included within the stacked image (Figure \ref{fig16}) are highlighted with dashed red vertical lines and correspond to periods of relatively low jet activity.}
\end{figure}

\begin{figure}[!htbp]
\begin{center}
\includegraphics[width=0.75\textwidth]{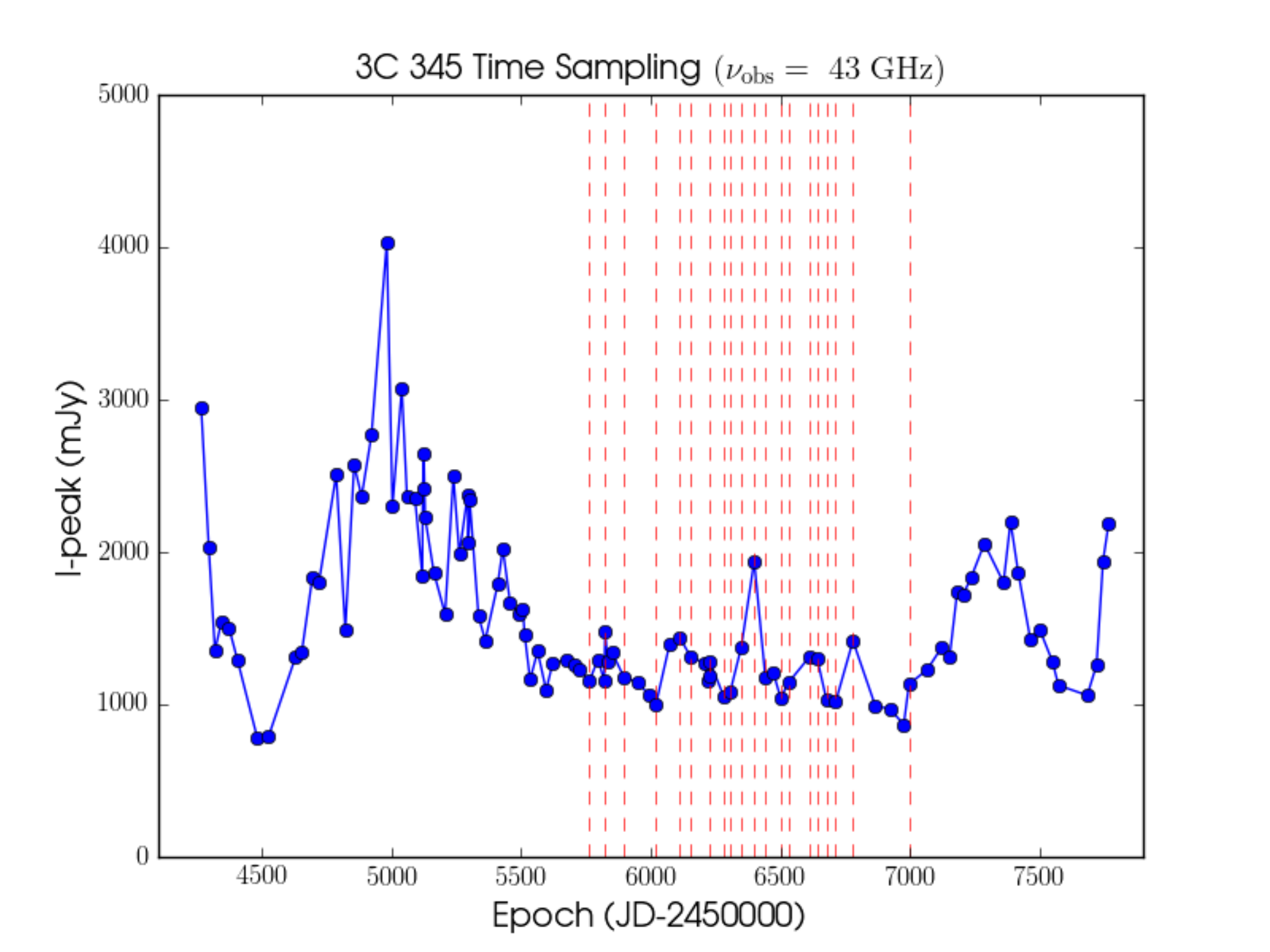}
\end{center}
\vspace{-0.3cm}
\caption{\label{fig27} Peak intensity at 43 GHz of 3C 345 versus time for the entire span of the VLBA-BU-BLAZAR program. The epochs included within the stacked image (Figure \ref{fig20}) are highlighted with dashed red vertical lines and correspond to periods of relatively low jet activity.}
\end{figure}

The selection of the epochs included in our stacking procedure have a significant impact of the final resultant images shown in Figures \ref{fig3}, \ref{fig6}, \ref{fig10}, \ref{fig13}, \ref{fig16}, \& \ref{fig20}. In this appendix we highlight the manner in which these epochs were selected. For each source in our sample we plot the peak intensity detected at 43 GHz, I-peak [mJy], versus the modified Julian date, (JD-2450000), of all the epochs within the VLBA-BU-BLAZAR campaign. The `quiescent' epochs included in each stacked map are highlighted with dashed vertical red lines and correspond to epochs of relatively low jet activity that still have detectable levels of polarization. This selection process is illustrated in Figures \ref{fig22}, \ref{fig23}, \ref{fig24}, \ref{fig25}, \ref{fig26}, \& \ref{fig27} for each of the blazars in our sample. 

\end{document}